\documentclass{apj}
\usepackage{CJKutf8}
\usepackage[caption=false]{subfig}
\usepackage{graphicx}
\usepackage{amsmath}
\usepackage{extarrows}
\usepackage{multirow}
\usepackage{graphicx}
\usepackage{booktabs}
\usepackage{amsmath,amssymb}
\usepackage[T1]{fontenc}
\graphicspath{{background/}}
\usepackage{longtable}

\usepackage{threeparttablex}

\usepackage{soul, color, xcolor}

\soulregister{\cite}7 % 注册\cite命令
\soulregister{\citep}7 % 注册\citep命令
\soulregister{\citet}7 % 注册\citet命令
\soulregister{\ref}7 % 注册\ref命令
\soulregister{\pageref}7 % 注册\pageref命令

\usepackage{ulem} % 导入 ulem 宏包

\begin{document}

\title{In-orbit background and sky survey simulation study of POLAR-2/LPD}

\correspondingauthor{Hong-Bang Liu}
\email{E-mail: liuhb@gxu.edu.cn}

\author{Zu-Ke Feng}
% \email{E-mail: 2207401001@st.gxu.edu.cn}
\affiliation{School of Physical Science and Technology, Guangxi University, Nanning 530004, China}
\author{Hong-Bang Liu}
\affiliation{School of Physical Science and Technology, Guangxi University, Nanning 530004, China}
\author{Fei Xie}
\affiliation{School of Physical Science and Technology, Guangxi University, Nanning 530004, China}
\author{Huan-Bo Feng}
\affiliation{School of Physical Science and Technology, Guangxi University, Nanning 530004, China}
\author{Qian-Nan Mai}
\affiliation{School of Physical Science and Technology, Guangxi University, Nanning 530004, China}
\author{Jiang-Chuan Tuo}
\affiliation{School of Physical Science and Technology, Guangxi University, Nanning 530004, China}
\author{Qian Zhong}
\affiliation{School of Physical Science and Technology, Guangxi University, Nanning 530004, China}
\author{Jian-Chao Sun}
\affiliation{Key Laboratory of Particle Astrophysics, Institute of High Energy Physics, Chinese Academy of Sciences, Beijing 100049, China}
\author{Jiang He}
\affiliation{Key Laboratory of Particle Astrophysics, Institute of High Energy Physics, Chinese Academy of Sciences, Beijing 100049, China}
\author{Yuan-Hao Wang}
\affiliation{Key Laboratory of Particle Astrophysics, Institute of High Energy Physics, Chinese Academy of Sciences, Beijing 100049, China}
\author{Qian Liu}
\affiliation{School of Physics Science, University of Chinese Academy of Sciences, Beijing, 100049, China}
\author{Di-Fan Yi}
\affiliation{School of Physics Science, University of Chinese Academy of Sciences, Beijing, 100049, China}
\author{Rui-Ting Ma}
\affiliation{School of Physics Science, University of Chinese Academy of Sciences, Beijing, 100049, China}
\author{Bin-Long Wang}
\affiliation{School of Physics Science, University of Chinese Academy of Sciences, Beijing, 100049, China}
\author{Zhen-Yu Tang}
\affiliation{Beijing Institute of Spacecraft Environment Engineering, Beijing 100094, China}
\author{Shuang-Nan Zhang}
\affiliation{Key Laboratory of Particle Astrophysics, Institute of High Energy Physics, Chinese Academy of Sciences, Beijing 100049, China}
\author{En-Wei Liang}
\affiliation{School of Physical Science and Technology, Guangxi University, Nanning 530004, China}

\begin{abstract}
% The Low-Energy X-ray Polarization Detector (LPD) is one of the three payloads in the POLAR-2, which is a successor of the POLAR experiment and will be deployed as an external payload on the China Space Station (CSS) in early 2026. LPD is dedicated to perform the polarization observation of Gamma-Ray Bursts (GRBs) prompt emission in 2-10 keV with a 90° wide field of view (FoV), using an array of X-ray photoelectric polarimeter based on the gas pixel detector. Under the scheme of the wide FoV, the in-orbit background count rate is high in the soft X-ray energy band even GRBs also have a high flux in this band. We simulate the main interactions in the instrument materials using the C++ package GEANT4 and present various kind of background components under the scheme of wide FoV. The simulation results show that among these background components, the main background components are cosmic X-ray background (CXB) and bright X-ray sources. The total background count rate of LPD, after applying charged particle background rejection algorithm, is approximately 0.6 counts/cm²/s. Detailed simulations and comparative analysis of CXB and X-ray bright sources are executed under different FoV sizes and pointings of the detector. 
The Low-Energy X-ray Polarization Detector (LPD) is one of the payloads in the POLAR-2 experiment, designed as an external payload for the China Space Station (CSS) deployment in early 2026. LPD is specifically designed to observe the polarization of Gamma-Ray Bursts (GRBs) prompt emission in the energy range of 2-10 keV, with a wide field of view (FoV) of 90 degrees in preliminary design. This observation is achieved using an array of X-ray photoelectric polarimeters based on gas pixel detectors.
Due to the wide FoV configuration, the in-orbit background count rate in the soft X-ray range is high, while GRBs themselves also exhibit a high flux in this energy band. In order to assess the contribution of various background components to the total count rate, we conducted detailed simulations using the GEANT4 C++ package. Our simulations encompassed the main interactions within the instrument materials and provided insights into various background components within the wide FoV scheme.
The simulation results reveal that among the background components, the primary contributors are the cosmic X-ray background (CXB) and bright X-ray sources. The total background count rate of LPD, after applying the charged particle background rejection algorithm, is approximately 0.55 counts/cm²/s on average, and it varies with the detector's orbit and pointing direction. Furthermore, we performed comprehensive simulations and comparative analyses of the CXB and X-ray bright sources under different FoVs and detector pointings. These analyses provide valuable insights into the background characteristics for soft X-ray polarimeter with wide FoV.
\end{abstract}


% \maketitle

%________________________________________________ sections below
%
\section{Introduction}           %% first-level sections will be auto-capitalized
\label{sect:intro}
Over the past few decades, there has been a growing interest in the detection of polarization in X-ray and gamma-ray astronomy, particularly for the study of one of the most powerful cosmic events: GRBs. Satellites like Swift \citep{burrows2005swift} and Fermi \citep{atwood2009large} have significantly advanced our understanding of GRBs by providing valuable observations of their energy spectra and timing properties, both during the prompt emission phase and the subsequent afterglow. Despite these advancements, several fundamental questions regarding the nature of GRBs remain unanswered, including the properties of the central engine, the mechanisms driving the energetic jets, the composition of the jets, the processes responsible for energy dissipation, the configurations of magnetic fields, and the mechanisms behind particle acceleration and radiation \citep{kumar2015physics,10.1111/j.1745-3933.2009.00624.x,10.1093/mnras/sts219,10.1111/j.1745-3933.2009.00624.x,10.1111/j.1365-2966.2010.17600.x,Zhang_2002,Bégué_2016,Rees_1994}.
\par
In contrast, the polarization of photons provides crucial insights into the magnetic field structures, angular structures of the jets, and emission mechanisms within the GRB emission region \citep{toma2009statistical,gill2020linear,galaxies9040082,10.1093/mnras/stab1013,10.1093/mnras/stu457,10.1111/j.1365-2966.2004.07387.x,Granot_2003}. By studying the polarization properties, we can gain a deeper understanding of the physical processes occurring in the vicinity of the central engine and the mechanisms responsible for the explosive release of energy during a GRB.

\par

The study of gamma-ray burst (GRB) polarization has been conducted using the principle of Compton scattering by several missions, including RHESSI (150 keV to 2 MeV) \citep{coburn2003polarization}, INTEGRAL-IBIS (250 keV to 800 keV) \citep{gotz2013polarized}, INTEGRAL-SPI (100 keV to 350 keV) \citep{kalemci2007search}, IKAROS-GAP (50 keV to 300 keV) \citep{yonetoku2012magnetic}, AstroSat-CZTI (100 keV to 300 keV) \citep{vadawale2015hard}, COSI (200 keV to 5 MeV) \citep{lowell2017polarimetric}, and POLAR (50 keV to 500 keV) \citep{zhang2019detailed}. Additionally, there are planned missions specifically designed for GRB detection, such as LEAP (30 keV to 500 keV) \citep{mcconnell2021large}. However, these instruments operate in the hard X-ray or gamma-ray range, leaving the soft X-ray polarization observation of GRBs largely unexplored.
\par
Theoretical studies suggest a correlation between the predicted polarization degree of different polarization models and the observed energy range. For instance, synchrotron radiation and Compton drag models predict lower polarization degrees at lower energy ranges \citep{toma2009statistical}, while the polarization from the photosphere model becomes more significant around several keV \citep{lundman2018polarization}. The Low-Energy X-ray Polarization Detector (LPD) aims to fill this gap by employing an array of photoelectric polarimeters with a wide FoV to achieve polarization detection of GRBs in the 2-10 keV energy range. Although projects like IXPE and eXTP also utilize similar techniques optimized for the soft energy range, they primarily focus on persistent point sources during long exposures and virtually impossible to detect GRB prompt emission, due to the narrow FoV caused by the focusing optics, and with slower reorientation. It is worth mentioning that, IXPE has reported polarimetric observations of the afterglow from the brightest GRB recorded to date, GRB221009A, providing valuable insights into its properties \citep{Negro_2023}. Additionally, due to the limited effective area of CubeSat platforms such as PolarLight, the number of collected photons is limited, making it challenging to achieve high-sensitivity polarization detection of GRBs.
\par

% The standard parameter to express the sensitivity to polarization is the Minimum Detectable Polarization (MDP) at a confidence level of 99\%.
The standard parameter used to quantify the sensitivity to polarization is the Minimum Detectable Polarization (MDP) at a confidence level of 99\%. It is defined as \citep{weisskopf2010understanding}:
\begin{equation}
M D P_{99}=\frac{4.29}{\mu S} \sqrt{\frac{S+B}{T}}
    \label{eq:MDP}
\end{equation}
% Where S and B are the count rate of the source and background, T is the total exposure time and $\mu$ is the modulation factor representing the amplitude of the response to 100\% polarized beam. Most GRBs have a high flux at the soft X-ray energy band, but the in-orbit background count rate also high in this energy range under the wide FoV. Understanding all kinds of background components is vital to fulfill polarization detection with high sensitivity. Taking these into account, a mass model of the LPD have been developed as shown in Fig. \ref{Massmodel} and the primary interactions in the instrument materials are simulated based on the Geant4 framework \citep{AGOSTINELLI2003250}. We analyze and simulate main background components deposit energy in sensitive regions of the working gas under the scheme of wide FoV. The simulation results show that among these background components, the main background sources are CXB and bright X-ray sources. We carry out detailed simulation and comparative analysis of CXB and X-ray bright sources under different FoV sizes and different pointings of the detector.
Where S and B are the count rate of the source and background, T is the total exposure time and $\mu$ is the modulation factor representing the amplitude of the response to 100\% polarized beam.
In the case of GRBs prompt emission, they often exhibit high flux in the soft X-ray range. However, it is crucial to understand and mitigate the background count rate, which is also high in this energy range due to the wide FoV of the instrument.
To address this, a detailed mass model of the LPD has been developed, as shown in Fig. \ref{Massmodel}, and simulations of the primary interactions within the instrument materials have been performed using the Geant4 framework \citep{AGOSTINELLI2003250,1610988}. These simulations allow for the analysis and characterization of the main background components that deposit energy in the sensitive regions of the working gas, considering the wide FoV scheme.
The simulation results highlight that the main background sources are the cosmic X-ray background (CXB) and bright X-ray sources. To gain a better understanding of these background components, we conducted detailed simulations and performed comparative analysis under different FoVs and different pointings of the detector.
\par

% \begin{table}
% \centering
% \caption{The main performance parameters of LPD.}
% \scalebox{0.9}{
% \begin{tabular}{cc}
% \hline
% \textbf{Parameters}                  & \textbf{Anticipated value}                        \\ \hline
% Energy range                         & 2-10 keV                                          \\
% Window                               & 50 $\mu \mathrm{m}$ Be                            \\
% Gas mixture                          & Ne 30\% $+$ DME 70\% at 0.8 atm           \\
% Absorption depth                     & 1.36 cm                                            \\
% Detection area                       & $\sim$ 298 $\mathrm{~cm}^2$                         \\
% FoV                        & $\sim 90^{\circ} \times 90^{\circ}$               \\
% Energy resolution (FWHM)             & $\leq 20$ \% @ 5.9 keV               
%    \\
% Minimum Detectable Polarization (MDP) & \textless 5\% @GRB190114C (vertical incidence)    \\ \hline
% \end{tabular}
% \label{performance}
% }
% \end{table}

\begin{table}[]
\centering
\begin{threeparttable}
\caption{The main performance parameters of LPD.}

\begin{tabular}{cc}
\hline
\textbf{Parameters}                  & \textbf{Anticipated value}                                                                                                                                                                  \\ \hline
Energy range                                                                                                      & 2-10 keV                                                                                                                                                                                                      \\
Window                                                                                                            & 50 $\mu \mathrm{m}$ Be                                                                                                                                                                                        \\
Gas mixture                                                                                                       & Ne 30\% $+$ DME 70\% at 0.8 atm                                                                                                                                                                               \\
Absorption depth                                                                                                  & 1.36 cm                                                                                                                                                                                                       \\
Detection area                                                                                                    & $\sim$ 298 $\mathrm{~cm}^2$                                                                                                                                                                                   \\
FoV                                                                                                               & $\sim 90^{\circ} \times 90^{\circ}$                                                                                                                                                                           \\
Energy resolution (FWHM)                                                                                          & $\leq 20$ \% @ 5.9 keV                                                                                                                                                                                        \\
\begin{tabular}[c]{@{}c@{}}Minimum Detectable Polarization (MDP)\\ for a 20-degree oblique incidence\end{tabular} & \begin{tabular}[c]{@{}c@{}}$\sim$ 7.8\% @GRB190114C\tnote{*}\\ $\sim$ 10.3\% @GRB191019A\\ $\sim$ 10.6\% @GRB210619B\\ $\sim$ 18.7\% @GRB201216C\\ $\sim$ 23.1\% @GRB210822A\\ $\sim$ 26.2\% @GRB210818A\\ $\sim$ 30.1\% @GRB201026A\end{tabular} \\ \hline
\end{tabular}
\begin{tablenotes}
        \footnotesize
        \item[*] The fluences of these GRBs at 15-150 keV are (data from official website of Swift/BAT\footnote{\href{https://www.swift.ac.uk/}{https://www.swift.ac.uk/}}, unit: $\operatorname{erg} / \mathrm{cm}^{2}$): $8.26 \times 10^{-5} \pm 9.57 \times 10^{-7}$ @GRB190114C, $1.04 \times 10^{-5} \pm 2.59 \times 10^{-7}$ @GRB191019A, $9.49 \times 10^{-5} \pm 1.07 \times 10^{-6}$ @GRB210619B, $4.51 \times 10^{-5} \pm 7.30 \times 10^{-7}$ @GRB201216C, $1.99 \times 10^{-5} \pm 3.84 \times 10^{-7}$ @GRB210822A, $2.62 \times 10^{-5} \pm 3.37 \times 10^{-7}$ @GRB210818A, $5.63 \times 10^{-6} \pm 2.43 \times 10^{-7}$ @GRB201026A.
  %此处加入注释*信息
      \end{tablenotes}
\end{threeparttable}
\label{performance}
\end{table}

This paper is organized as follows: In Section \ref{sec:POLAR-2/LPD}, we provide an introduction to the POLAR-2 mission and the LPD payload. 
Section \ref{sec:Background} focuses on the background characterization. We present the input spectrum of various background components and describe the simulated results. Additionally, we discuss the algorithm used for background rejection, specifically targeting charged particles. 
In Section \ref{sect:Src}, we delve into the X-ray sources catalogue and simulated sky survey. We describe the methodology and techniques employed to create the catalogue, which includes a comprehensive survey of X-ray sources. This section provides essential information about the sources considered in the analysis and their impact on the overall observations.
Finally, Section \ref{sect:summary} presents a summary of this work and a comprehensive discussion.
% We first introduce the POLAR-2 mission as well as LPD payload in Section \ref{sec:POLAR-2/LPD}. In Section \ref{sec:Background}, we present the input spectrum and simulated results of various background components as well as algorithm of background rejection for charged particles. The X-ray sources catalogue and simulated sky survey is described In Section \ref{sect:Src}. The summary and discussion are given in Section \ref{sect:summary}.

\begin{figure*}
\centering
\includegraphics[scale=0.5]{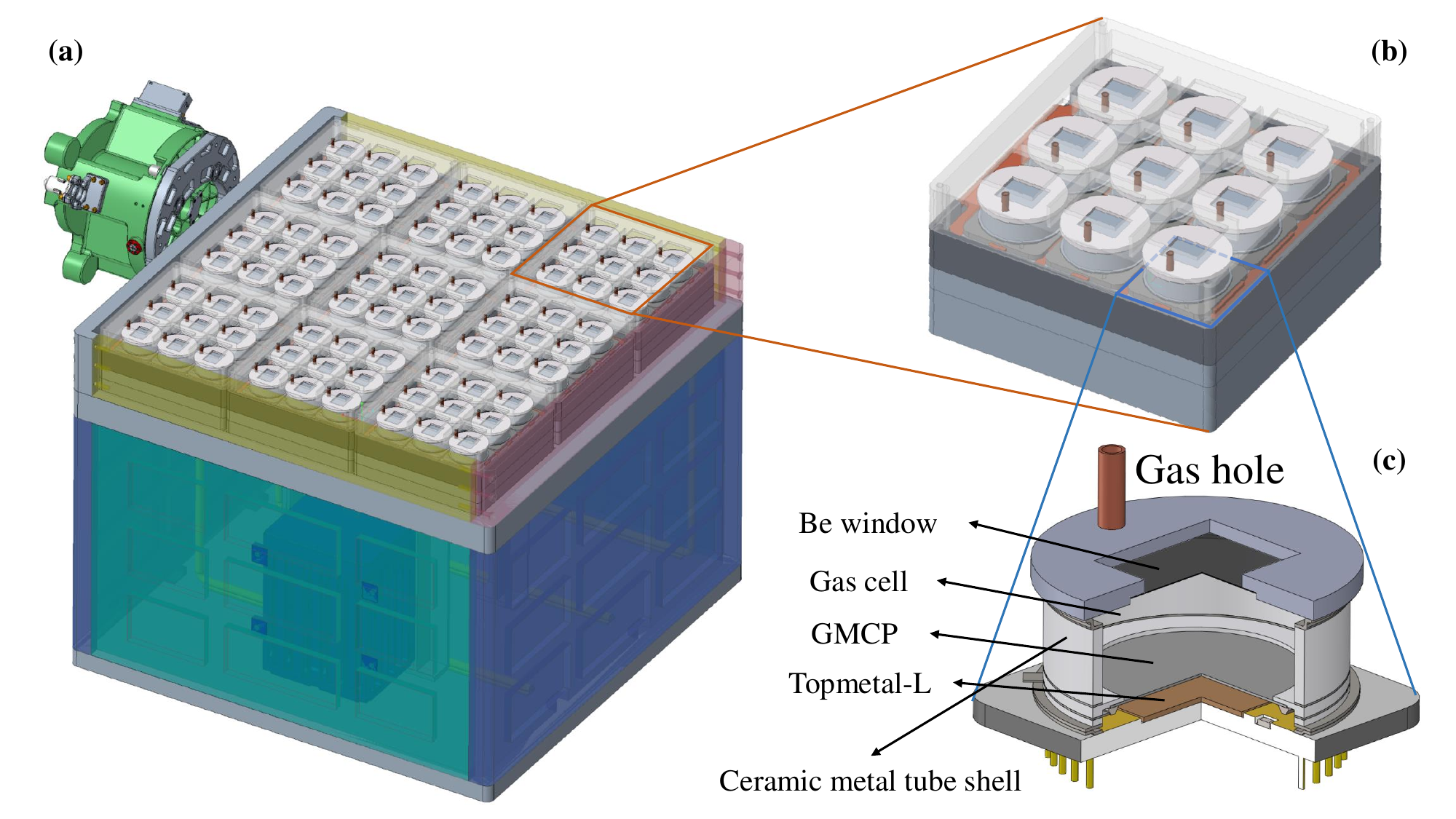}
\caption{Mechanical structures of LPD. (a) The LPD payload. (b) The detector array module. Each module includes a 3x3 array of detector units. (c) The detector unit. The main components of the gas chamber have been indicated in the diagram. Additionally, the material of the top cover is Kovar, the material of the gas hole is copper, the chamber walls are made of alumina ceramic, and the bottom consists of a ceramic base.
}
\label{Figure1}
\end{figure*}

\begin{figure*}
\centering
\includegraphics[scale=0.17]{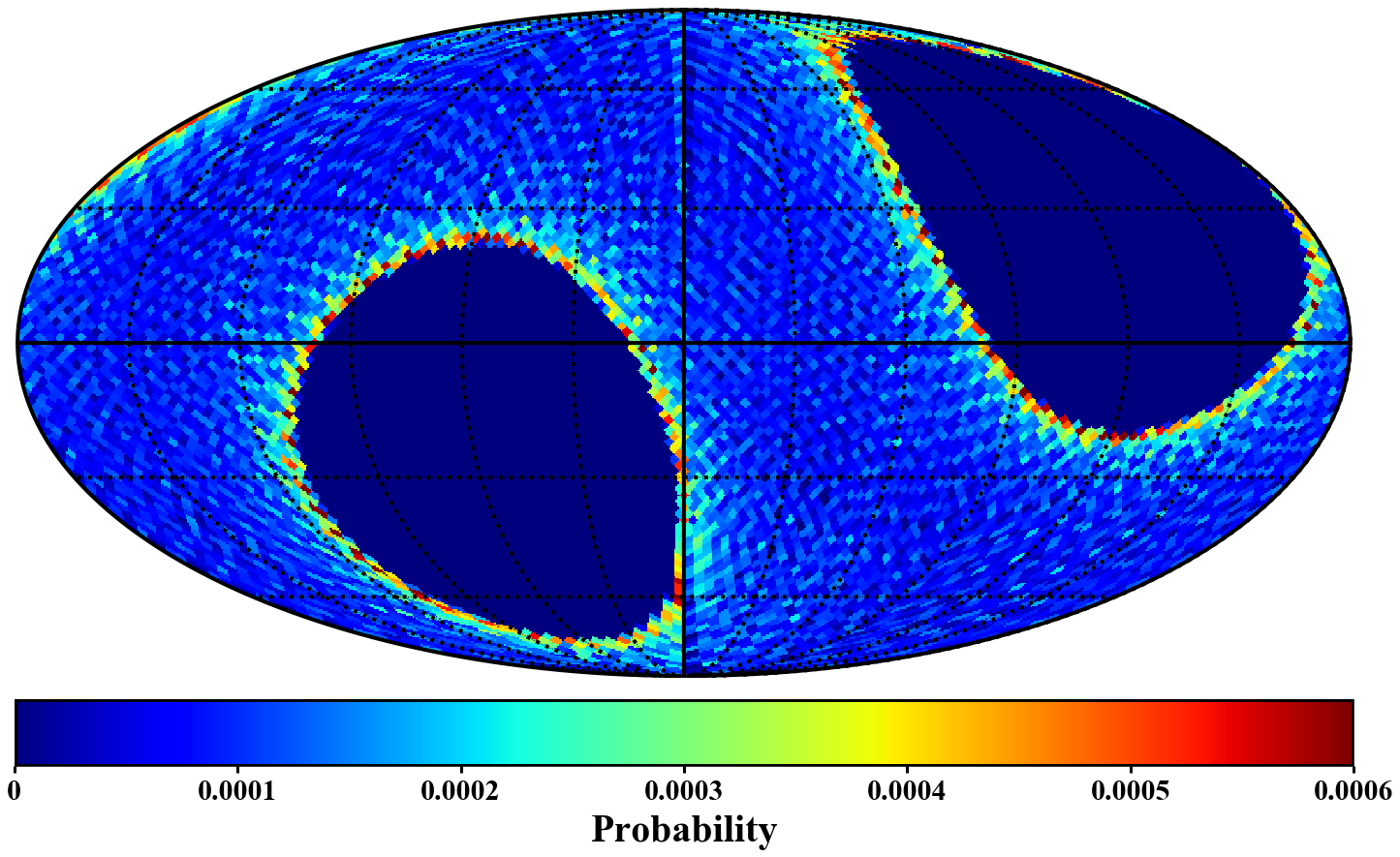}
\includegraphics[scale=0.17]{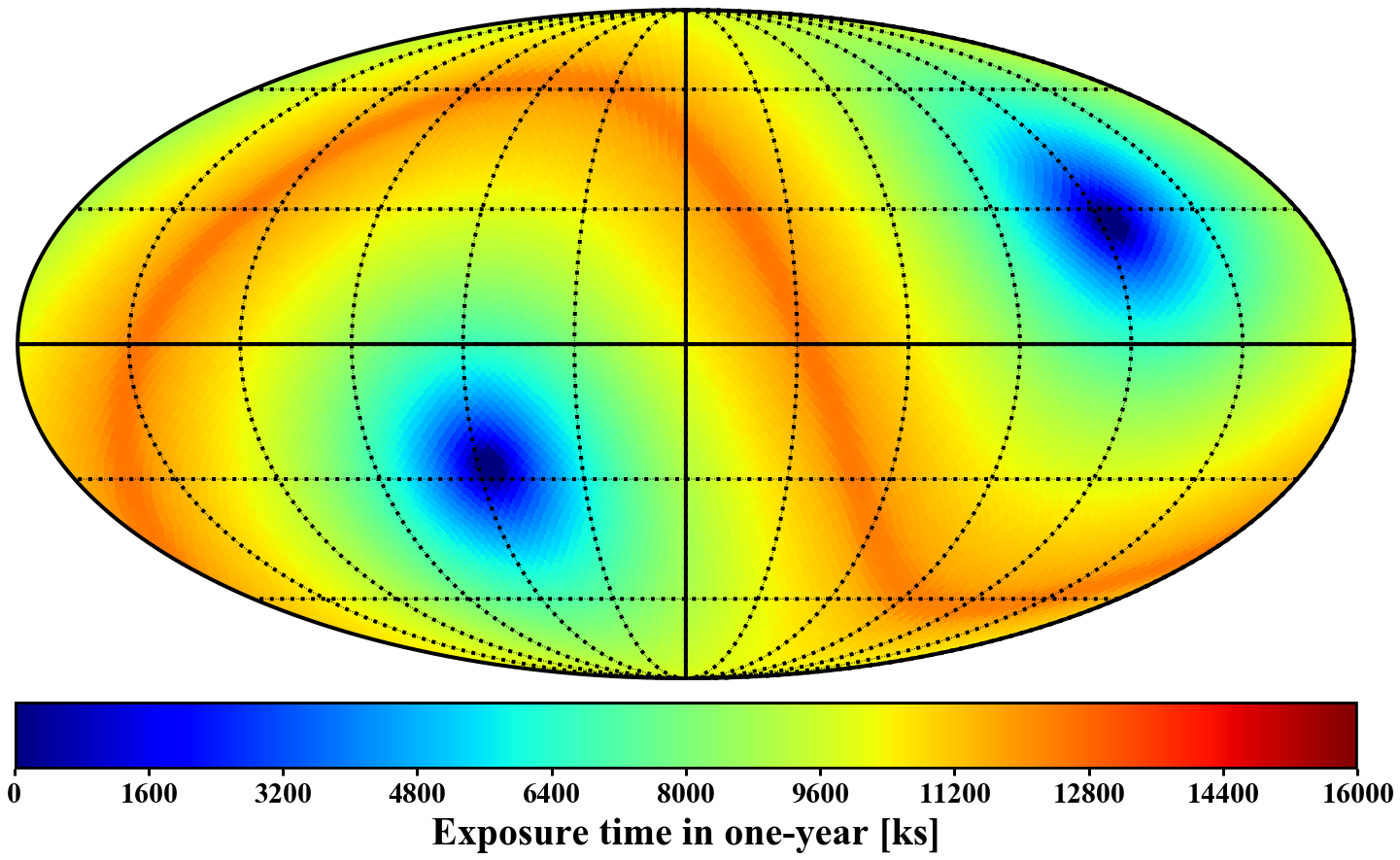}
\caption{Left panel: Distribution of pointing directions in Galactic coordinates during a one-year simulation. The color indicates the normalized probability of pointing in each direction (total 12288 pixels). Right panel: Distribution of exposure time during a one-year simulation.}
\label{pointsim}
\end{figure*}

\section{The POLAR-2 mission and LPD}
\label{sec:POLAR-2/LPD}
% POLAR-2 is a successor of the POLAR \citep{PRODUIT2018259} experiment. It will be deployed as an external payload on the China Space Station in 2026 and will operate for at least 2 years. POLAR-2 consists of a high-energy polarization detector (HPD, 30-800 keV, Europe)  \citep{hulsman2020polar}, a low-energy polarization detector (LPD, 2-10 keV, China) and a broad-spectrum detector (BSD, 8-2000 keV, China). In addition to the significant improvement in the detection ability of high-energy polarization, it also has capabilities of soft X-ray polarization measurement, wide energy spectrum measurement, timing measurement and location. 
POLAR-2 is the successor of the POLAR experiment \citep{PRODUIT2018259} and will be deployed as an external payload on the CSS in 2026. It consists of three detectors: a high-energy polarization detector (HPD) covering 30-800 keV developed by Europe \citep{hulsman2020polar}, a low-energy polarization detector (LPD) covering 2-10 keV developed by China, and a broad-spectrum detector (BSD) covering 8-2000 keV developed by China. In addition to the significant improvement in the detection ability of high-energy polarization, it also has capabilities of soft X-ray polarization measurement, wide energy spectrum measurement, timing measurement and location.

\begin{figure}
\centering
\includegraphics[scale=0.6]{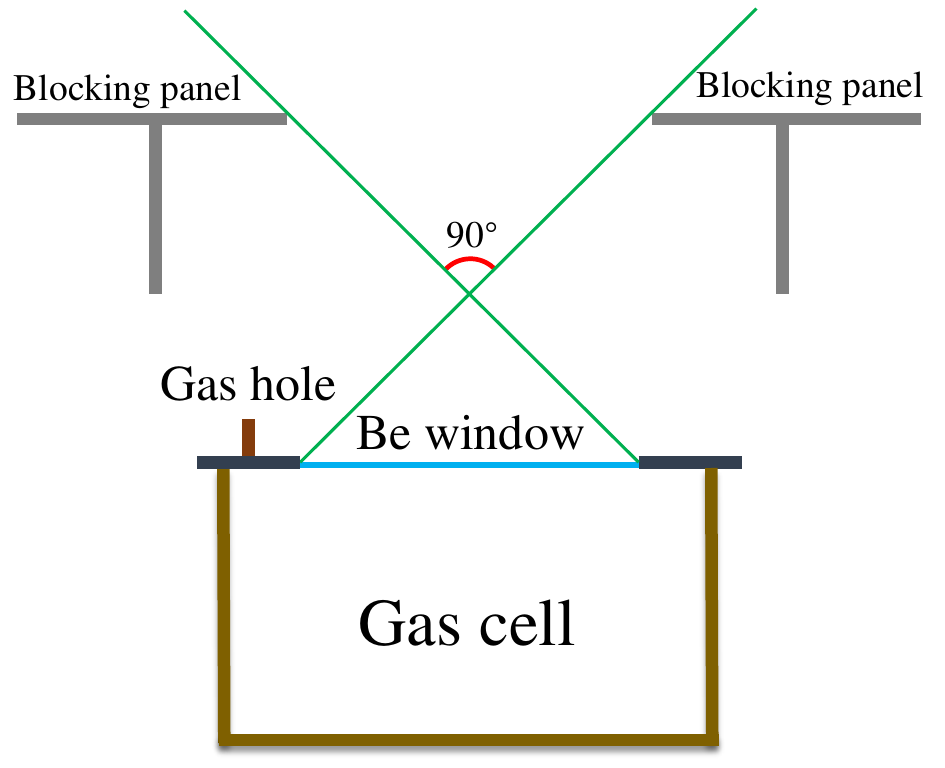}
\caption{Definition of FoV. The diagram provides an example of defining the FoV size using a 90-degree FoV, as indicated by the green line.}
\label{FoV}
\end{figure}
\begin{figure}
\centering
\includegraphics[scale=0.5]{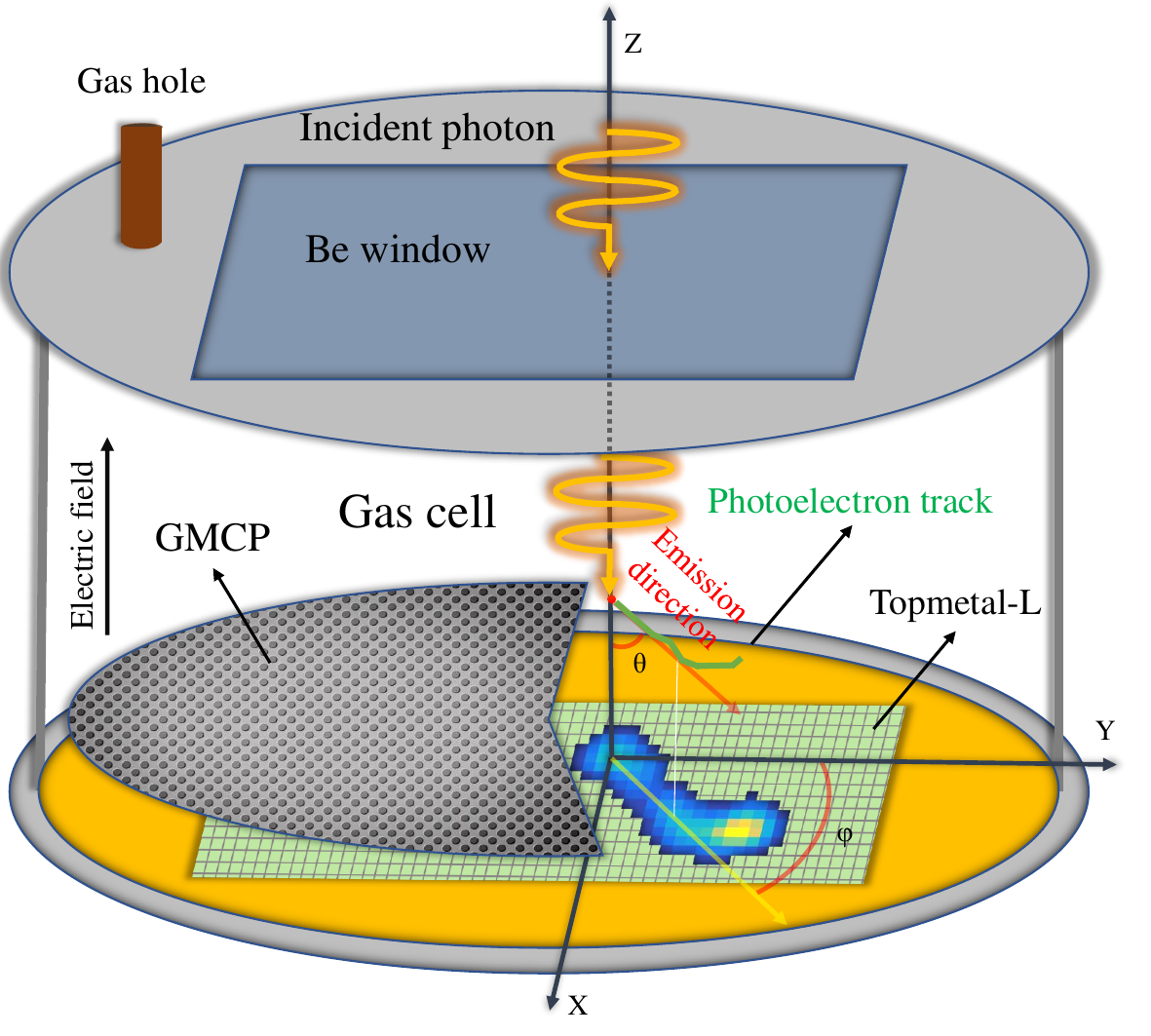}
\caption{Polarization detection  principle of microchannel plate pixel detector. When an X-ray photon penetrates the detector through the beryllium window and interacts with the working gas within the GMPD, photoelectron are emitted. It ionize the gas molecules, generating primary electrons and ions. Guided by the electric field, the primary electrons drift towards the GMCP holes, undergoing an avalanche amplification as they move towards the anode. While some electrons migrate to the top of the silicon pixel chip Topmetal-L, others are collected by the electrodes at the bottom of the GMCP. By reconstructing the emission direction of the photoelectron tracks on Topmetal-L, the polarization of the incident X-ray can be determined. Additionally, information about the X-ray's energy and time can be extracted by measuring the electrons collected at the bottom of the GMCP.}
\label{Detection_principle}
\end{figure}

% \vspace{-5em}

\par

% \subsection{Low-Energy X-ray Polarization Detector - LPD}
% LPD is a new scheme of wide FoV soft X-ray polarization detector using Gas Microchannel Plate pixel Detector (GMPD, a photoelectric X-ray polarimeter) array proposed by Guangxi University in China. The main parameters and expected performances of LPD are listed in Table \ref{performance}.
LPD is an innovative soft X-ray polarization detector with a wide FoV (the definition of FoV is shown in Fig. \ref{FoV}), utilizing the Gas Microchannel Plate pixel Detector (GMPD) array \citep{FENG2023168499,Feng_2023,2023ExA...tmp...46X}. It was proposed by Guangxi University in China. The performance and key parameters of LPD are summarized in Table \ref{performance}, showcasing its anticipated capabilities and characteristics.
\par
\subsection{Geometric structure}
% LPD consists of 15 detector modules arranged in a $3 \times 5$ detector array. Each detector modules contain 6 detector units, total 90 detector units. Each units is composed of cathode, GMCP, pixel readout chip and gas sealed chamber as shown in Fig. \ref{Figure1}(c). The GMCP is used as the electron multiplying device, and the Topmetal-L produced by the CMOS process is used as the readout chip to directly measure the track of the photoelectrons, thereby derive the polarization information of the incident X-ray. Topmetal-L is an updated version of Topmetal-II- \citep{li2021preliminary, fan2023front}, designed specifically to detect transient source for LPD, with a smaller pixel size and larger chip area. The LPD unit structure is mainly composed of a gas microchannel plate pixel detector and an electronic system. The detector is used to measure the photoelectron track, and the readout electronics system is to control the detector to read, organize and package the data. The ceramic metal sealed chamber of LPD is made of beryllium window, support frame, ceramic metal tube shell and ceramic base sealed by laser welding. The GMCP and Topmetal-L chip are packaged in the gas chamber. The copper intake tube is use to fill working gas, and finally is sealed by ultrasonic welding.
LPD is comprised of a total of 9 detector modules arranged in a $3 \times 3$ array configuration. Each detector module contains 9 detector units, resulting in a total of 81 detector units. Each unit consists of a cathode, Gas Microchannel Plate (GMCP), pixel readout chip (Topmetal-L), and a gas-sealed chamber, as illustrated in Fig. \ref{Figure1} (c). The GMCP serves as the electron multiplying device, while the Topmetal-L, manufactured using the CMOS process, acts as the readout chip responsible for measuring the tracks of photoelectrons and deriving the polarization information of the incident X-ray. Topmetal-L represents an upgraded version of Topmetal-II \citep{li2021preliminary, fan2023front}, specifically designed for LPD to detect transient sources. It features a smaller pixel size and a larger chip area, measuring 16mm $\times$ 23mm. The LPD unit structure mainly consists of a gas microchannel plate pixel detector and an electronic system. The detector is utilized to measure the photoelectron track, while the readout electronics system controls the detector's data reading, organization, and packaging. The ceramic-metal sealed chamber of LPD is constructed using a beryllium window, support frame, ceramic-metal tube shell, and ceramic base, all sealed through laser welding. The GMCP and Topmetal-L chip are packaged within the gas chamber. The copper intake tube is used for filling the working gas, which is then sealed using ultrasonic welding.
\par
\subsection{Working principle}
% The polarization detection principle of microchannel plate pixel detector is the same as typical GPD \citep{costa2001efficient} utilizing photoelectric effect as shown in Fig. \ref{Detection_principle}. When an X-ray photon enters the detector through the beryllium window and is absorbed by the working gas of the GMPD, a photoelectron are emitted. The photoelectrons ionize the gas molecules, resulting in the production of primary electrons and ions. The primary electrons subsequently drift into the hole of the Gas MicroChannel Plate (GMCP) under the influence of the electric field, where they undergo an avalanche and drift towards the anode. Some of these electrons migrate to the top of the silicon pixel chip Topmetal-L, while others are collected by the electrodes at the bottom of the GMCP. The X-ray polarization can be determined by reconstructing the emission direction of the photoelectron tracks on Topmetal-L. The energy and time of the X-rays can be obtained by measuring the electrons on the bottom of the GMCP.
The microchannel plate pixel detector operates on the same polarization detection principle as the conventional Gas Pixel Detector (GPD), as illustrated in Fig. \ref{Detection_principle} \citep{costa2001efficient}. When an X-ray photon penetrates the detector through the beryllium window and interacts with the working gas within the GMPD, photoelectron are emitted. It ionize the gas molecules, generating primary electrons and ions. Guided by the electric field, the primary electrons drift towards the GMCP holes, undergoing an avalanche amplification as they move towards the anode. While some electrons migrate to the top of the silicon pixel chip Topmetal-L, others are collected by the electrodes at the bottom of the GMCP. By reconstructing the emission direction of the photoelectron tracks on Topmetal-L, the polarization of the incident X-ray can be determined. Additionally, information about the X-ray's energy and time can be extracted by measuring the electrons collected at the bottom of the GMCP.

% \begin{figure}
% \centering
% \includegraphics[scale=0.42]{Effective_area.pdf}
% \caption{The effective area at different oblique incidence angles of LPD.}
% \label{Effective_area}
% \end{figure}

% \begin{figure}
% \centering
% \includegraphics[scale=0.19]{modulation curve.pdf}
% \caption{Modulation factor as a function of photon energy in neon (30\%) mixed with dimethyl ether (DME70\%) at 0.8 atmosphere. The photons are 100\% polarized and enter the detector vertically. We use the first moment to reconstruct the emission direction of the photoelectron for all energy points, in order to obtain a high modulation factor at low energy.}
% \label{Modulation_curve}
% \end{figure}

\section{Background studies}
\label{sec:Background}
\begin{figure*}
\centering
\includegraphics[scale=0.2]{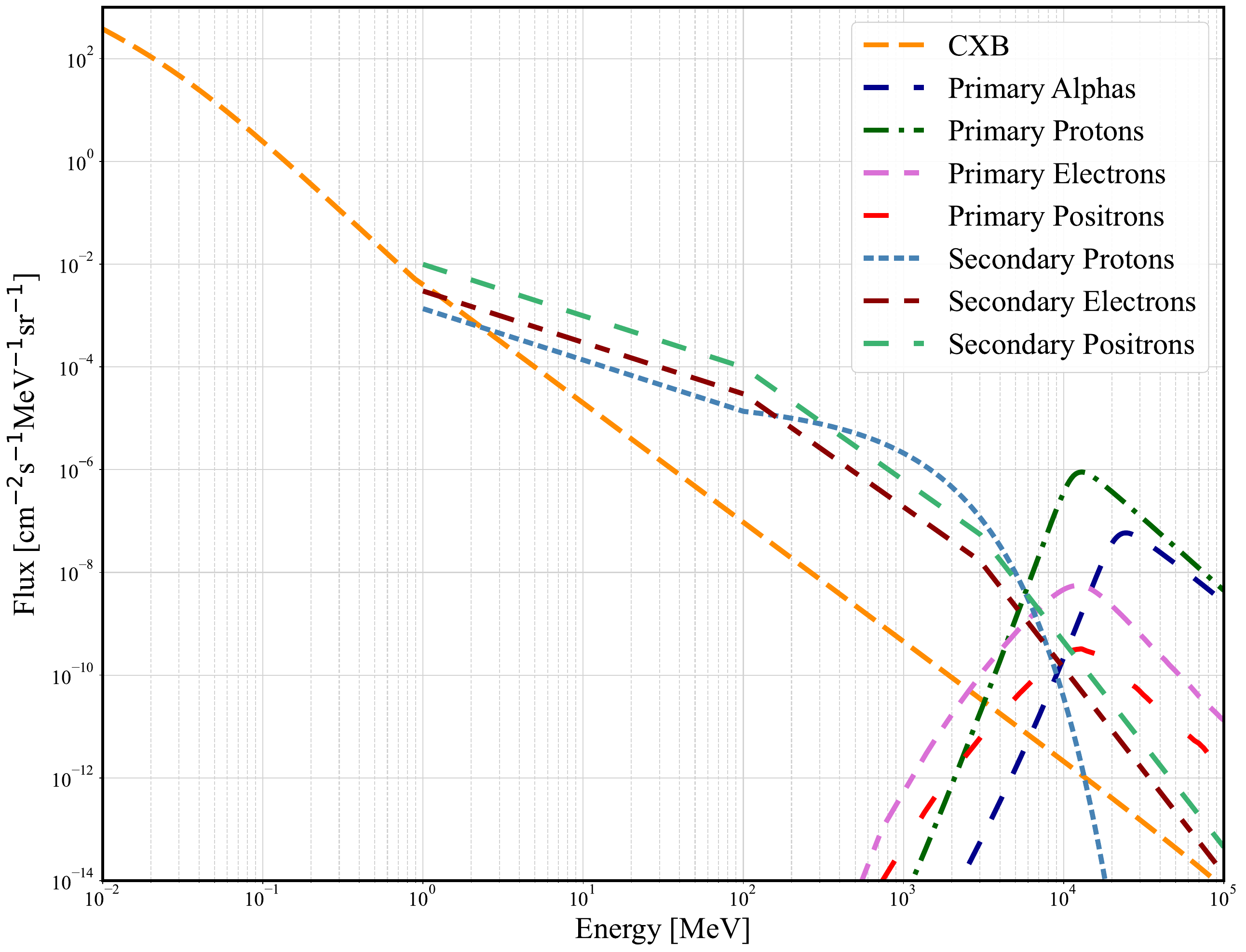}
\includegraphics[scale=0.2]{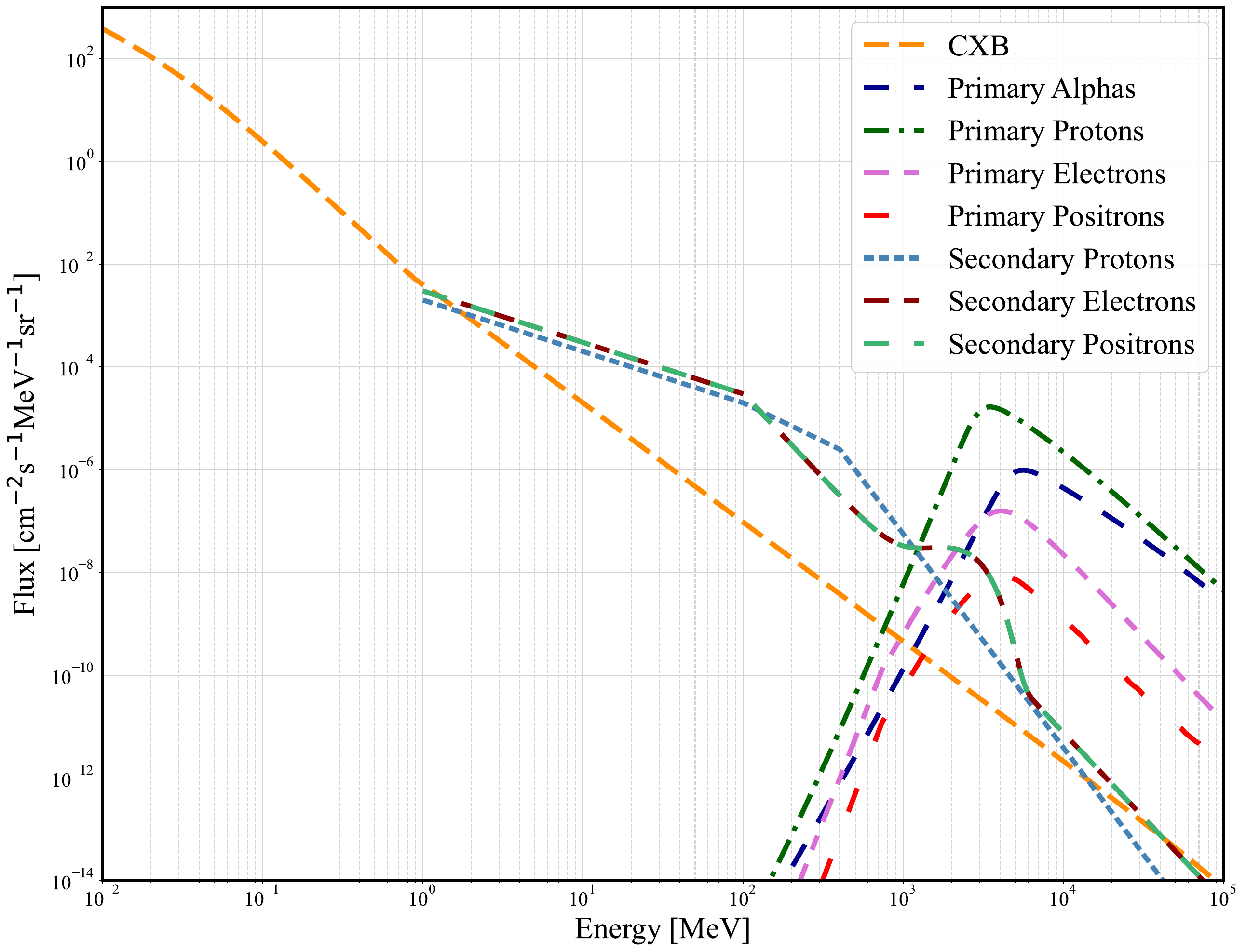}
\caption{The spectra of main background components expected in the CSS orbital space environment at two magnetic latitudes. (Left panel) 0°. (Right panel) 42°. As shown in the figure, the spectrum of the CXB does not vary with changes in geomagnetic latitude. Additionally, due to the different geomagnetic cut-off rigidities at various geomagnetic latitudes, the cut-off energy of primary charged particles varies significantly.}
\label{basp}
\end{figure*}

% LPD is expected to be mounted as an external payload on the China space station and always pointing away from the Earth. The orbital altitude of the China space station is approximately 400 km, with an inclination of approximately 42°. The background simulation results of IXPE \citep{xie2021study} and the in-orbit measurement results of Polarlight \citep{huang2021modeling} indicate that for gas photoelectric X-ray polarimeters with focusing telescopes or narrow FoV size, the primary background are charged particles, including electrons and protons. However, due to the wide FoV design of LPD, the background photons entering the FoV  will significantly increase, mainly including cosmic
% X-ray background and X-ray sources with known positions. Considering the expected flight orbit and direction away from Earth of LPD, the in orbit backgrounds under the wide FoV scheme mainly includes the following components: CXB, primary and secondary charged particles, charged particles in the South Atlantic anomaly and X-ray sources with known position. 
LPD is expected to be mounted as an external payload on the CSS, always pointing away from the Earth. The orbital altitude of the CSS is approximately 400 km, with an inclination of approximately 42°. Background simulation results from IXPE \citep{xie2021study} and in-orbit measurement results from Polarlight \citep{huang2021modeling} indicate that gas photoelectric X-ray polarimeters with focusing telescopes or narrow FoV primarily experience background noise from charged particles, including electrons and protons. However, due to LPD's wide FoV design, the number of background photons entering the FoV will significantly increase, primarily comprising cosmic X-ray background and X-ray sources with known positions. Taking into account LPD's expected flight orbit and direction away from Earth, the in-orbit backgrounds under the wide FoV scheme mainly consist of the following components: CXB, primary and secondary charged particles, charged particles in the South Atlantic anomaly, and X-ray sources with known positions.

\subsection{Space Radiation Environment}
% The input source energy spectrum models of main background components diffused in space acquired from the measurements of other missions as shown in Fig. \ref{basp}. The figure shows the background energy spectra at two extreme geomagnetic latitudes under the LPD orbit \citep{2019ExA....47..273C}. It include CXB, primary and secondary charged particles. We did not consider the albedo components due to the pointing strategy of POLAR-2.
The input energy spectrum models of the main background components were obtained from measurements conducted by other missions, as shown in Fig. \ref{basp}. This figure presents the background energy spectra at two extreme geomagnetic latitudes corresponding to the LPD orbit \citep{2019ExA....47..273C}. These spectra include the contributions from CXB, primary and secondary charged particles. Albedo components were not considered in our analysis due to the pointing strategy of POLAR-2.
\begin{figure*}
\centering
\includegraphics[scale=0.125]{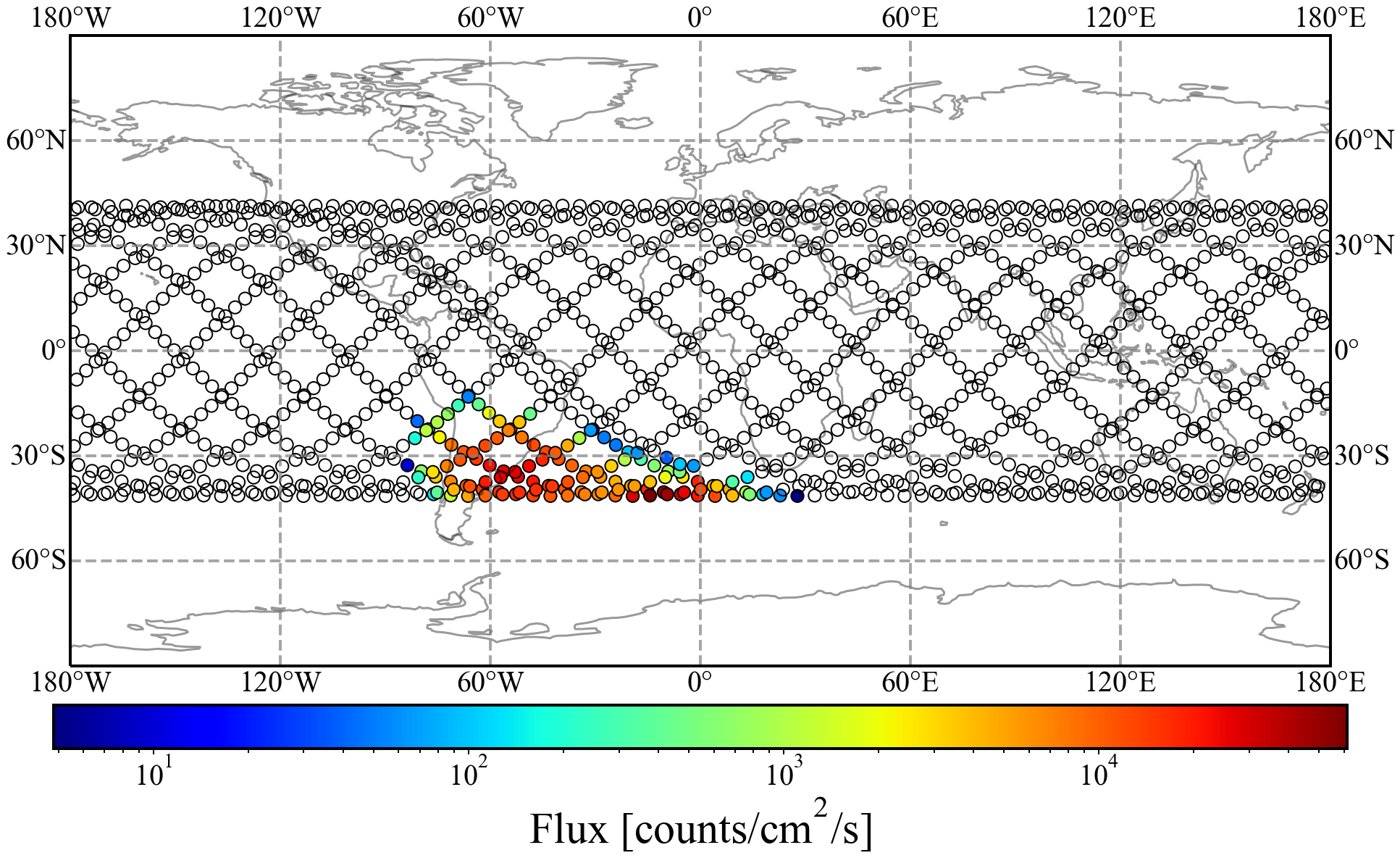}
\includegraphics[scale=0.125]{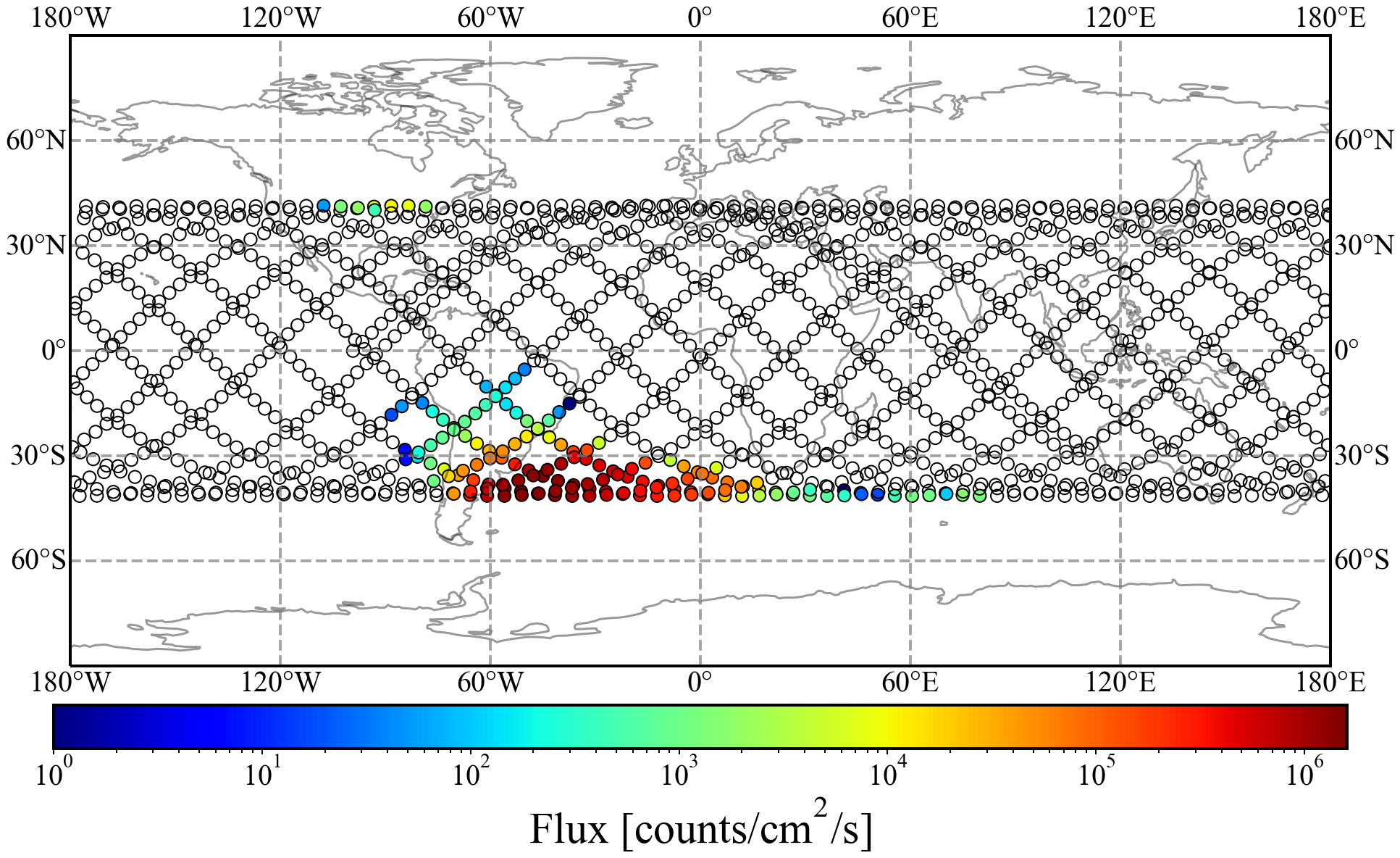}
\caption{The orbit of the CSS in one day and  the flux map of high energy (> 100 keV) charged particles in South Atlantic anomaly and polar regions (solar minimum). Left panel: Proton flux. Right panel: Electron flux (data generated by SPENVIS).}
\label{SAA}
\end{figure*}

\subsubsection{Cosmic X-ray background} 

% It has been more than 50 years since the Cosmic X-ray background was discovered. At present, it is believed that the cosmic X-ray originates from the contribution of a large number of extragalactic X-ray sources. In the soft X-ray range, it mainly comes from the contribution of active galactic nuclei, and some point sources in the hard X-ray range are still undetermined. Their intensity basically does not change with time, and the spatial distribution is isotropic \citep{dean2003modelling}. The spectrum of CXB can be described as a broken power-law distribution \citep{gehrels1992instrumental} by the Eq. (\ref{E1}):
It has been over 50 years since the discovery of the Cosmic X-ray background \citep{PhysRevLett.9.439}. Currently, it is understood that the cosmic X-ray emission is attributed to the combined contribution of numerous extragalactic X-ray sources. In the soft X-ray range, the dominant contribution comes from active galactic nuclei, while the origin of some point sources in the hard X-ray range remains uncertain \citep{Giacconi_2001}. The intensity of CXB remains relatively constant over time, and the spatial distribution is isotropic \citep{dean2003modelling}. The spectrum of the Cosmic X-ray background can be described by a broken power-law distribution \citep{gehrels1992instrumental} as given in Eq. (\ref{E1}):
\begin{equation}
\label{E1}
\frac{\mathrm{d} N_\mathrm{C X B}}{\mathrm{~d} E}= \begin{cases}0.54 E^{-1.4} & E<0.02 \mathrm{MeV} \\ 0.0117 E^{-2.38} & 0.02 \leq E<0.1 \mathrm{MeV} \\ 0.014 E^{-2.3} & E \geq 0.1 \mathrm{MeV}\end{cases}
\end{equation}
where the $\mathrm{d} N_\mathrm{C X B} / \mathrm{d} E$ is in units of counts $\mathrm{cm}^{-2}$ $\mathrm{s}^{-1}$ $\mathrm{MeV}^{-1}$ $\mathrm{sr}^{-1}$.

\begin{figure}
\centering
\includegraphics[scale=0.50]{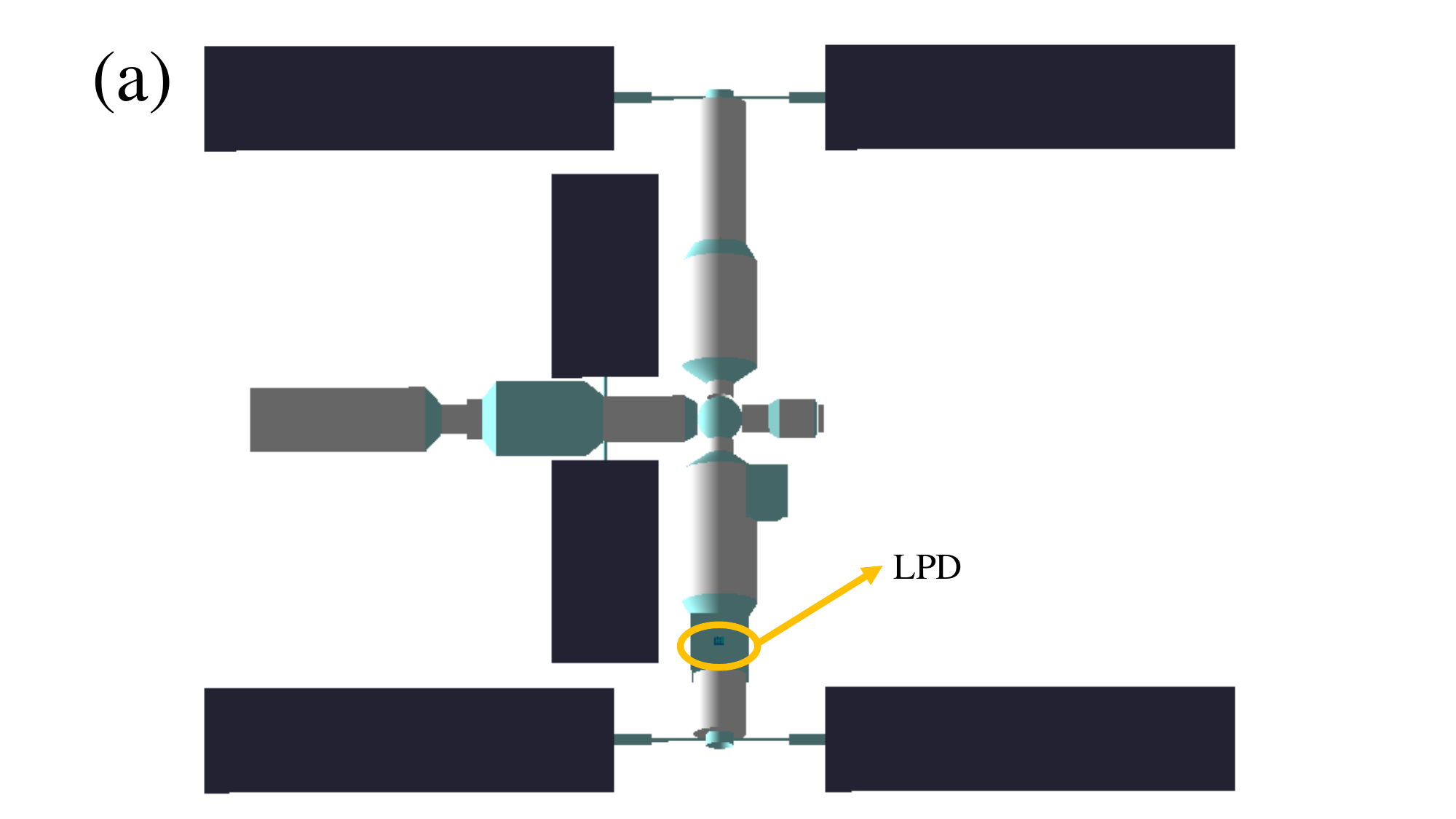}
\includegraphics[scale=0.50]{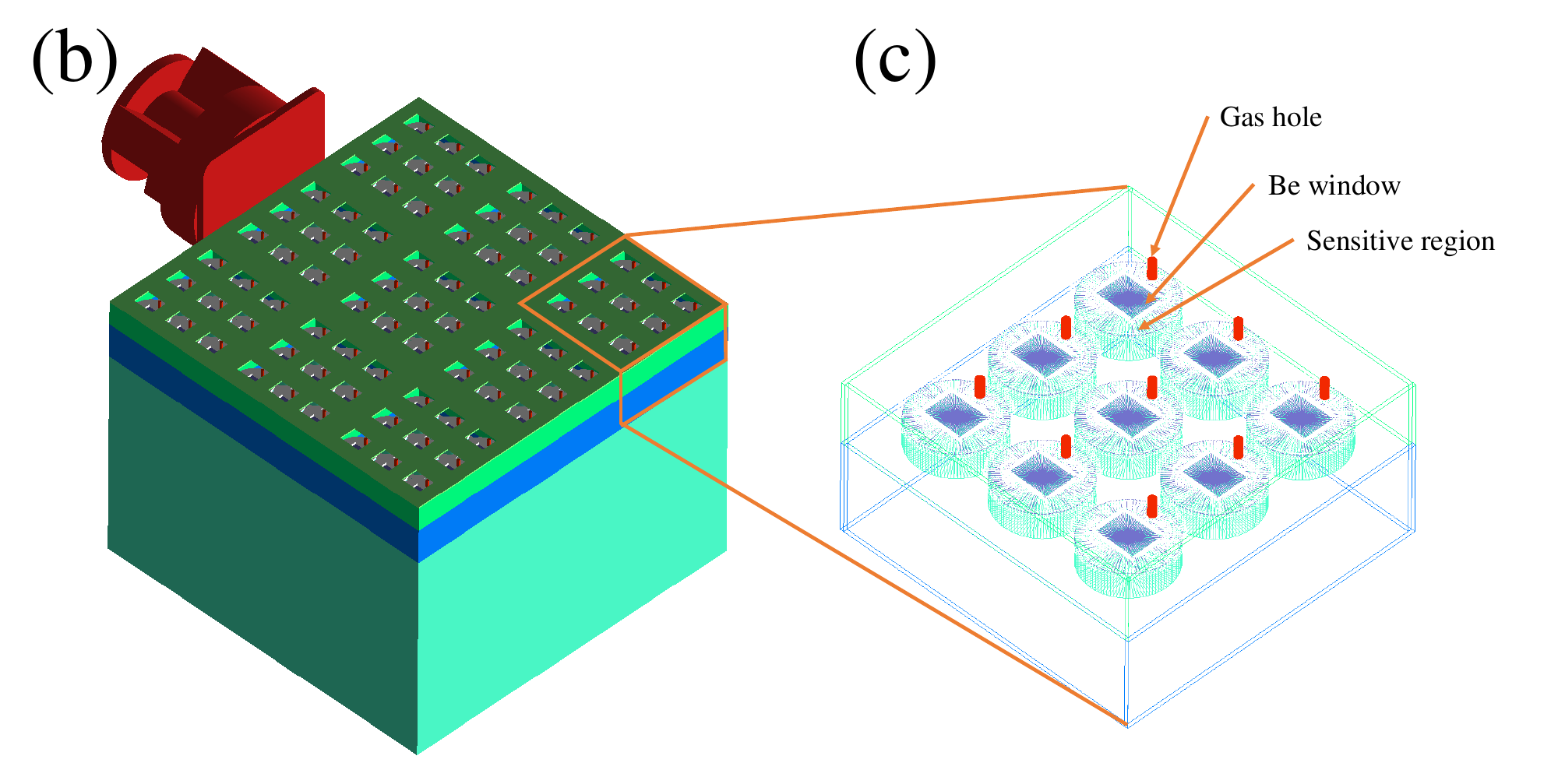}
\caption{The LPD and CSS mass model was built using Geant4. (a) The simplified model of the CSS features aluminium alloy as its primary construction material. The anticipated position of the LPD is indicated by a yellow circle. (b) The LPD payload, including the mechanical arm adapter (red, main material is aluminium alloy), and the LPD detector array (green). The FoV material is lead glass. (c) The detector module, each of which includes a 3x3 arrangement of gas chambers. The top cover of the gas chamber is made of Kovar alloy, and a copper gas tube traverses through the top cover to establish a connection between the gas and the external environment. The upper end of the copper tube is sealed. The material of the side wall of the gas chamber is aluminum oxide ceramic, which is supported by a ceramic base underneath. At the bottom of the detection gas defines a simplified unperforated GMCP, made of lead glass, with a thickness of 400 micrometers. The upper surface of the GMCP features a layer of plated metal, composed of NiCr alloy, with a thickness of 0.1 micrometers.}
\label{Massmodel}
\end{figure}

\begin{figure}
\centering
\includegraphics[scale=0.4]{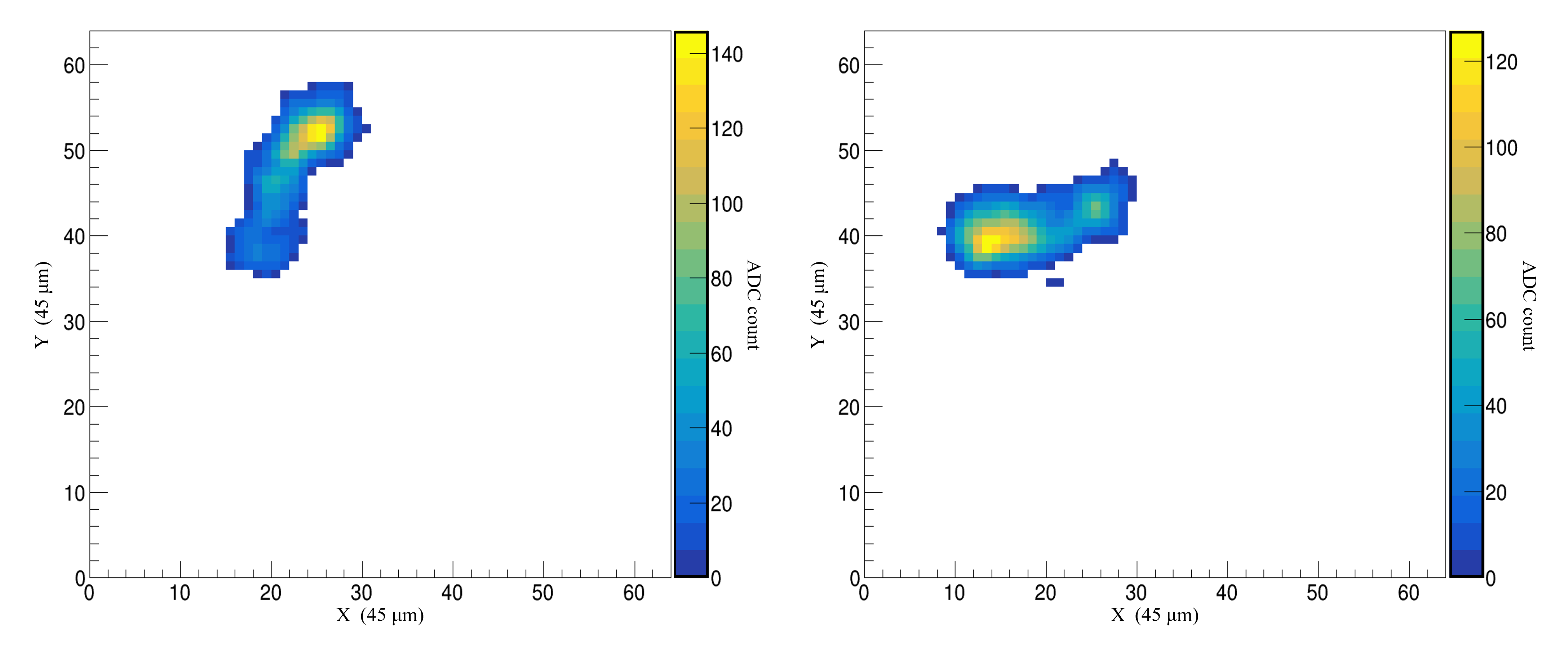}
\caption{The simulated photoelectron track generated by incident photons with an energy of 6 keV. The detection gas used in the simulation consists of a mixture of Ne (30\%) and DME (70\%) at a pressure of 0.8 atmosphere. The side length of each square pixel in the image is 45 micrometers. The two endpoints of the track image correspond to Auger electron and Bragg peak (higher energy deposition).}
\label{Photoelectron_track}
\end{figure}
\subsubsection{Cosmic rays}

% Cosmic rays in low-Earth orbit are mainly composed of primary and secondary components. Primary cosmic rays come from outside the solar system and are mainly composed of protons. Secondary cosmic ray particles come from the product of the interaction between the original composition and the atmosphere. It's energy is relatively low relative to the original composition, mainly concentrated below GeV. When the primary particles approach the earth, they will be affected by the modulation of solar activity (mainly affecting the low-energy charged particles below GeV) and the geomagnetic field. To enter the geomagnetic field, the magnetic rigidity of the charged particles must be greater than the geomagnetic cut-off rigidity. By fitting the observation data of AMS \citep{alcaraz2000leptons,alcaraz2000cosmic} at different geomagnetic latitudes and different observation directions from 1998 to 1999, the primary cosmic ray energy spectrum of low-Earth orbit was obtained. The secondary cosmic ray proton spectrum is also obtained by fitting the AMS observational data. The primary component spectrum is fitted by a Cutoff power law spectrum, and the secondary component is fitted by a Broken power law spectrum \citep{mizuno2004cosmic}. 

Cosmic rays in low-Earth orbit consist primarily of two components: primary and secondary cosmic rays. Primary cosmic rays originate from outside the solar system and are predominantly composed of protons. Secondary cosmic rays are the product of interactions between the primary cosmic rays and the Earth's atmosphere. Their energies are relatively lower compared to the primary component, mainly concentrated below GeV. As the primary particles approach the Earth, they are influenced by solar activity modulation (which mainly affects low-energy charged particles below GeV) and the Earth's geomagnetic field.

To enter the geomagnetic field, the charged particles must have a magnetic rigidity greater than the geomagnetic cut-off rigidity. By analyzing the observation data from the Alpha Magnetic Spectrometer (AMS) at different geomagnetic latitudes and observation directions from 1998 to 1999, the energy spectrum of primary cosmic rays in low-Earth orbit was derived \citep{alcaraz2000leptons,alcaraz2000cosmic}. The spectrum of secondary cosmic ray protons was also obtained through fitting of the AMS observational data. The primary component spectrum was fitted with a cutoff power-law distribution, while the secondary component was fitted with a broken power-law distribution \citep{mizuno2004cosmic}.
\subsubsection{SAA} 
% The center of the SAA area is located at 45° west longitude and 30° south latitude, and the overall scale extends from 15° east longitude to 120° west longitude. The overall magnetic field intensity in the area is low, about one-half of normal, so the radiation flux of charged particles is very large. LPD will shut down to protect the machine when it passes by. What's more, it will activate some elements in the instrument, and elements with long half-lives will still contribute to the background after normal power-on. 
% \par
% The orbit of the China space station and the proton/electron flux in South Atlantic anomaly as well as polar regions are shown in Fig. \ref{SAA} (data generated by SPENVIS\footnote{\href{https://www.spenvis.oma.be/}{https://www.spenvis.oma.be/}}). About 8\% times in one day passes through the South Atlantic anomaly and reaches the poles according to the results.
The South Atlantic anomaly (SAA) is centered at 45° west longitude and 30° south latitude, with its overall extent ranging from 15° east longitude to 120° west longitude. In this region, the overall intensity of the Earth's magnetic field is low, approximately half of the normal intensity, resulting in a high flux of radiation from charged particles. To protect the LPD instrument from this intense radiation, it will be shut down when passing through the SAA. However, even after normal power-on, the trapped high energy protons in the SAA region will activate some elements. Some nuclides in the instrument with long half-lives will continue to contribute to the background.

Fig. \ref{SAA} shows the orbit of the CSS and the proton/electron flux in the South Atlantic anomaly and polar regions (data generated by SPENVIS\footnote{\href{https://www.spenvis.oma.be/}{https://www.spenvis.oma.be/}}). According to the results, LPD is estimated to pass through the South Atlantic anomaly and the poles approximately 8\% of the time in one day.

\subsection{LPD massmodel and simulation logic}
% We perform in orbit background response studies through Monte Carlo simulation. The mass model of LPD built by Geant4\footnote{\href{http://geant4.web.cern.ch/}{http://geant4.web.cern.ch/}} simulation toolkit are shown in Fig. \ref{Massmodel}. It Contains the main materials of the LPD payload, omitting some details, such as screws. The Geant4 package version 4.10.07 is utilized for mass modeling and simulation. The Livermore models in Geant4 are adopted for low energy electromagnetic processes. X-ray polarization , Auger electrons and radioactive decay are enabled. For the photoelectron track simulation, the gas above the chip in the mass model is set as a sensitive volume, and Geant4 is used to simulate and track every step information of the interaction of photoelectrons in gas. After obtaining the simulated track backbone, each step goes through the digitization process, where the transverse Gaussian diffusion, proportional to the square root of the drift distance to the upper surface of GMCP, and charge multiplication of GMCP are applied. The parameters for diffusion and multiplication are meticulously chosen to replicate the characteristics of the real experimental process to the greatest extent possible. Finally, the diffused and multiplied photoelectron track is projected into each pixel of the Topmetal-L chip to form the track as shown in Fig. \ref{Photoelectron_track}. The track reconstruction algorithm study has been discussed in \citep{huang2021simulation} for LPD.\par

We conduct in-orbit background response studies using Monte Carlo simulation. The mass model of LPD is created using the Geant4\footnote{\href{http://geant4.web.cern.ch/}{http://geant4.web.cern.ch/}} simulation toolkit, as shown in Fig. \ref{Massmodel}. This model includes the main materials of the LPD payload while omitting certain details like screws. We utilize Geant4 package version 4.10.07 for mass modeling and simulation. The Livermore models in Geant4 are employed for low-energy electromagnetic processes, and features such as X-ray polarization, Auger electrons, and radioactive decay are enabled.
In the simulation of photoelectron tracks, the gas above the chip in the mass model is designated as a sensitive volume. Geant4 is used to simulate and track every step of the interaction between photoelectrons and the gas. Once the simulated track backbone is obtained, each step undergoes a digitization process. This process includes transverse Gaussian diffusion, which is proportional to the square root of the drift distance to the upper surface of the GMCP, and charge multiplication of the GMCP. We meticulously choose parameters for diffusion and multiplication to replicate the characteristics of the real experimental process as closely as possible.
Finally, the diffused and multiplied photoelectron track is projected onto each pixel of the Topmetal-L chip to form the track, as depicted in Fig. \ref{Photoelectron_track}. The track reconstruction algorithm for LPD has been discussed in detail in a previous study \citep{huang2021simulation}.

% For background simulation, the CXB and primary charged particle background are assumed to have a uniform spherical incidence over a 4π solid angle. Additionally, the shielding effect of the Earth needs to be considered, resulting in an effective solid angle of 8.3 sr \citep{2019ExA....47..273C}. As for the secondary charged particle background, it is divided into two components: vertical downward and vertical upward. Each component is assumed to have a uniform spherical incidence over a 2π solid angle. In the case of low geomagnetic latitudes, the flux and energy spectra of these two components are approximately equal. For charged particles in SAA are assumed to have a uniform spherical incidence over a 4π solid angle.

For background simulation, we assume that the Cosmic X-ray Background (CXB) and primary charged particle background have a uniform spherical incidence over a 4$\Pi$ sr. We also take into account the shielding effect of the Earth, which results in an effective solid angle of 8.3 sr \citep{2019ExA....47..273C} in 400 km orbital altitude.
Regarding the secondary charged particle background, it is divided into two components: downward and upward. Each component is assumed to have a uniform spherical incidence over a 2$\Pi$ sr. At low geomagnetic latitudes, the flux and energy spectra of these two components are approximately equal. The solar modulation potential is set at 650 MV.
As for the charged particles in the South Atlantic Anomaly (SAA), we assume that they have a uniform spherical incidence over a 4$\Pi$ sr.

\begin{table*}
% \centering
\caption{Simulation results of energy deposition in the 2-10 keV range at 0 magnetic latitude for different background components, considering the presence or absence of the CSS (unit: $\operatorname{counts} / \mathrm{cm}^{2} / \mathrm{s}$).}
 \scalebox{0.67}{
\begin{tabular}{@{}ccccccccc@{}}
\toprule
\begin{tabular}[c]{@{}c@{}}Particle\\ type\end{tabular} & \begin{tabular}[c]{@{}c@{}}Priamry\\ Alpha\end{tabular} & \begin{tabular}[c]{@{}c@{}}Priamry\\ Proton\end{tabular} & \begin{tabular}[c]{@{}c@{}}Priamry\\ Electron\end{tabular} & \begin{tabular}[c]{@{}c@{}}Priamry\\ Positron\end{tabular} & \begin{tabular}[c]{@{}c@{}}Secondary\\ Proton\end{tabular} & \begin{tabular}[c]{@{}c@{}}Secondary\\ Electron\end{tabular} & \begin{tabular}[c]{@{}c@{}}Secondary\\ Positron\end{tabular} & CXB      \\ \midrule
 CSS presence                                               & 4.77E-03                                                & 4.21E-02                                                 & 3.64E-04                                                   & 2.20E-05                                                   & 3.10E-02                                                   & 4.56E-02                                                     & 1.69E-01                                                     & 4.51E-01 \\
CSS absence                                                  & 5.50E-03                                                & 5.42E-02                                                 & 3.96E-04                                                   & 2.50E-05                                                   & 5.01E-02                                                   & 8.23E-02                                                     & 2.64E-01                                                     & 4.57E-01 \\ \bottomrule
\end{tabular}
}
\label{spacestation}
\end{table*}

\begin{table*}
% \centering
\caption{Simulation results of charged particles depositing energy in 2-10 keV at different magnetic latitudes (unit: $\operatorname{counts} / \mathrm{cm}^{2} / \mathrm{s}$).}
 \scalebox{1.0}{
\begin{tabular}{lcccccc}
\hline
Magnetic Latitude  & 0°       & 10°      & 20°      & 30°      & 40°      & 42°      \\ \hline
Priamry Alpha      & 5.50E-03 & 6.21E-03 & 8.38E-03 & 1.39E-02 & 2.68E-02 & 3.15E-02 \\
Priamry Proton     & 5.42E-02 & 6.04E-02 & 8.32E-02 & 1.40E-01 & 2.89E-01 & 3.34E-01 \\
Priamry Electron   & 3.96E-04 & 4.58E-04 & 6.82E-04 & 1.31E-03 & 3.23E-03 & 3.96E-03 \\
Priamry Positron   & 2.50E-05 & 2.87E-05 & 4.04E-05 & 7.58E-05 & 1.84E-04 & 2.28E-04 \\
Secondary Proton   & 5.01E-02 & 5.15E-02 & 1.91E-02 & 1.59E-02 & 2.02E-02 & 2.74E-02 \\
Secondary Electron  & 8.23E-02 & 8.20E-02 & 9.56E-03 & 7.62E-02 & 7.31E-02 & 7.24E-02 \\
Secondary Positron & 2.64E-01 & 2.74E-01 & 1.28E-01 & 1.25E-01 & 7.33E-02 & 7.19E-02 \\
Total              & 4.57E-01 & 4.75E-01 & 2.49E-01 & 3.72E-01 & 4.86E-01 & 5.41E-01 \\ \hline
\end{tabular}
}
\label{background_resp}
\end{table*}

\subsection{Simulation results}
\label{sect: simresult}

\begin{figure*}
\centering
\includegraphics[scale=0.4]{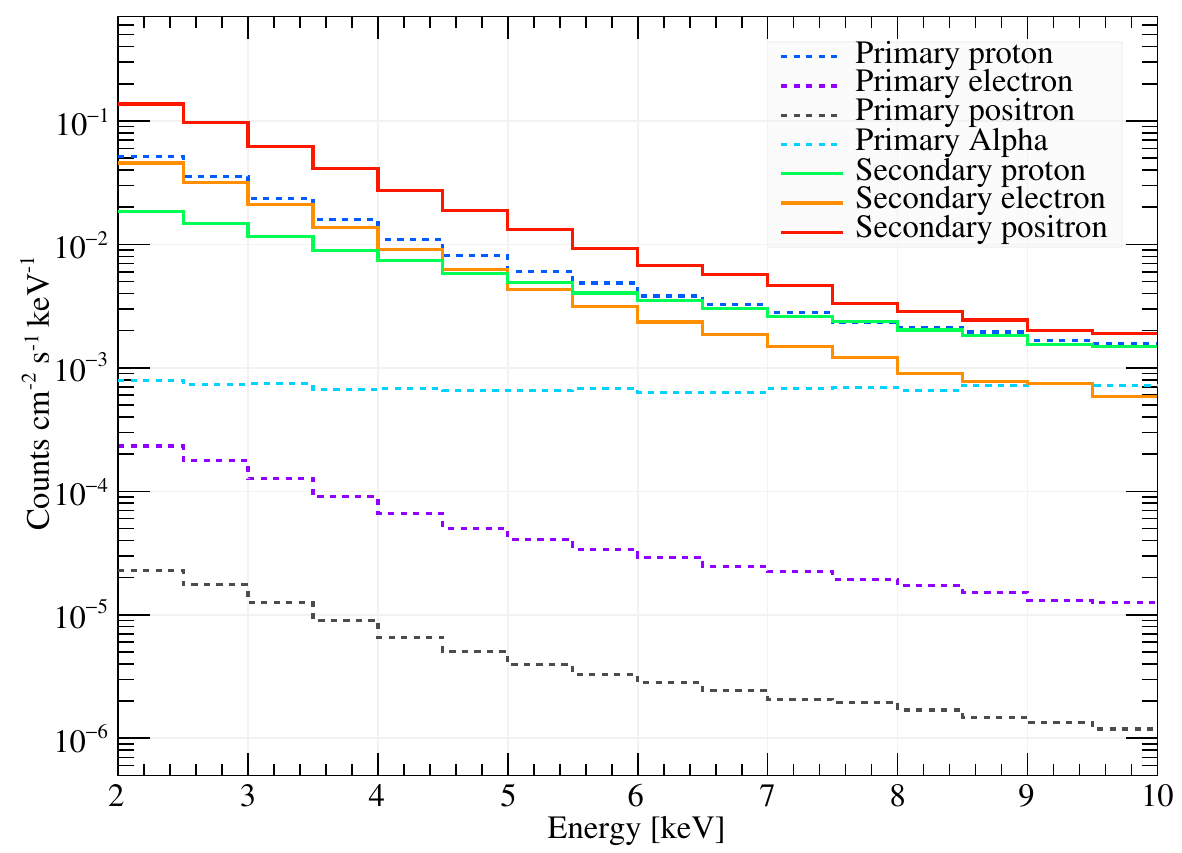}
\includegraphics[scale=0.4]{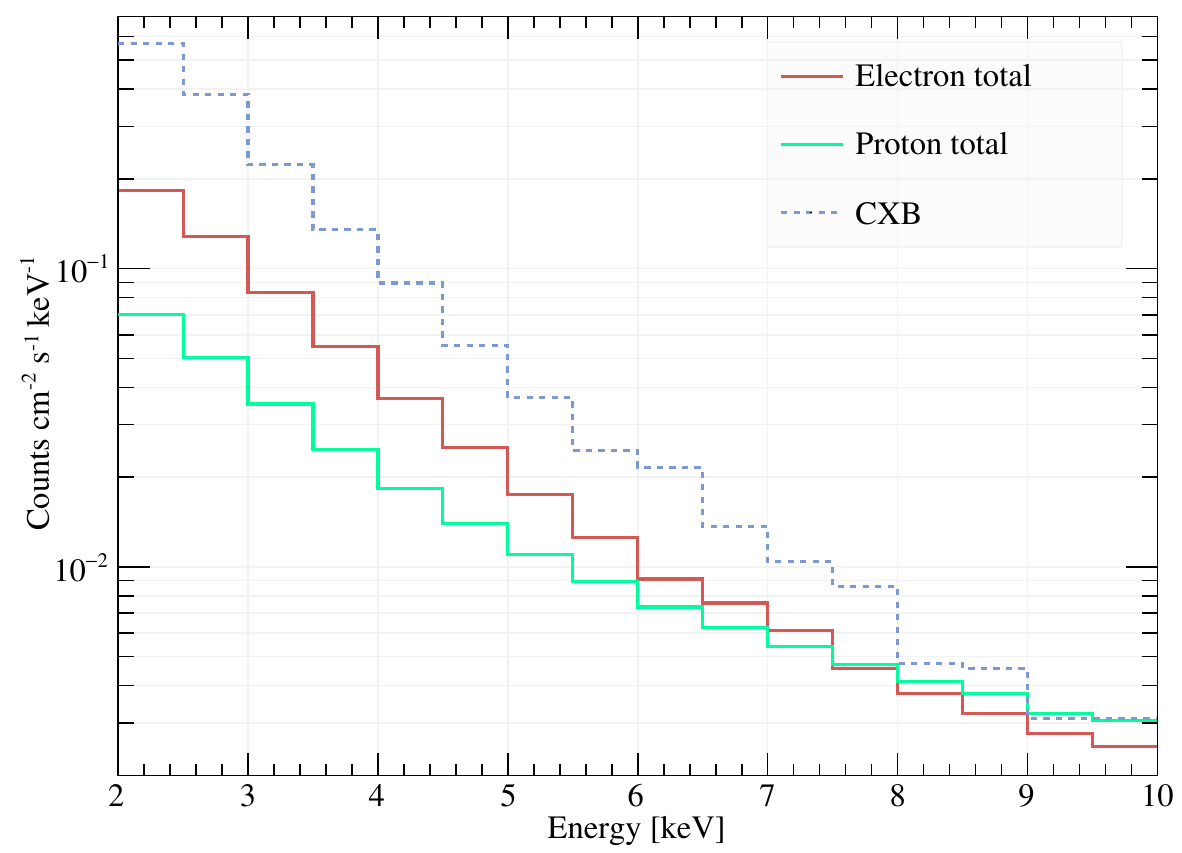}
\caption{Simulated spectral responses of different background components.  Left panel: The energy spectrum response of various charged particle backgrounds. Right panel: The total energy spectrum response of electrons, protons, and CXB background is presented. It should be noted that the category of electrons includes both positrons and electrons.}
\label{Gas_Edep}
\end{figure*}

\begin{figure}
\centering
\includegraphics[scale=0.30]{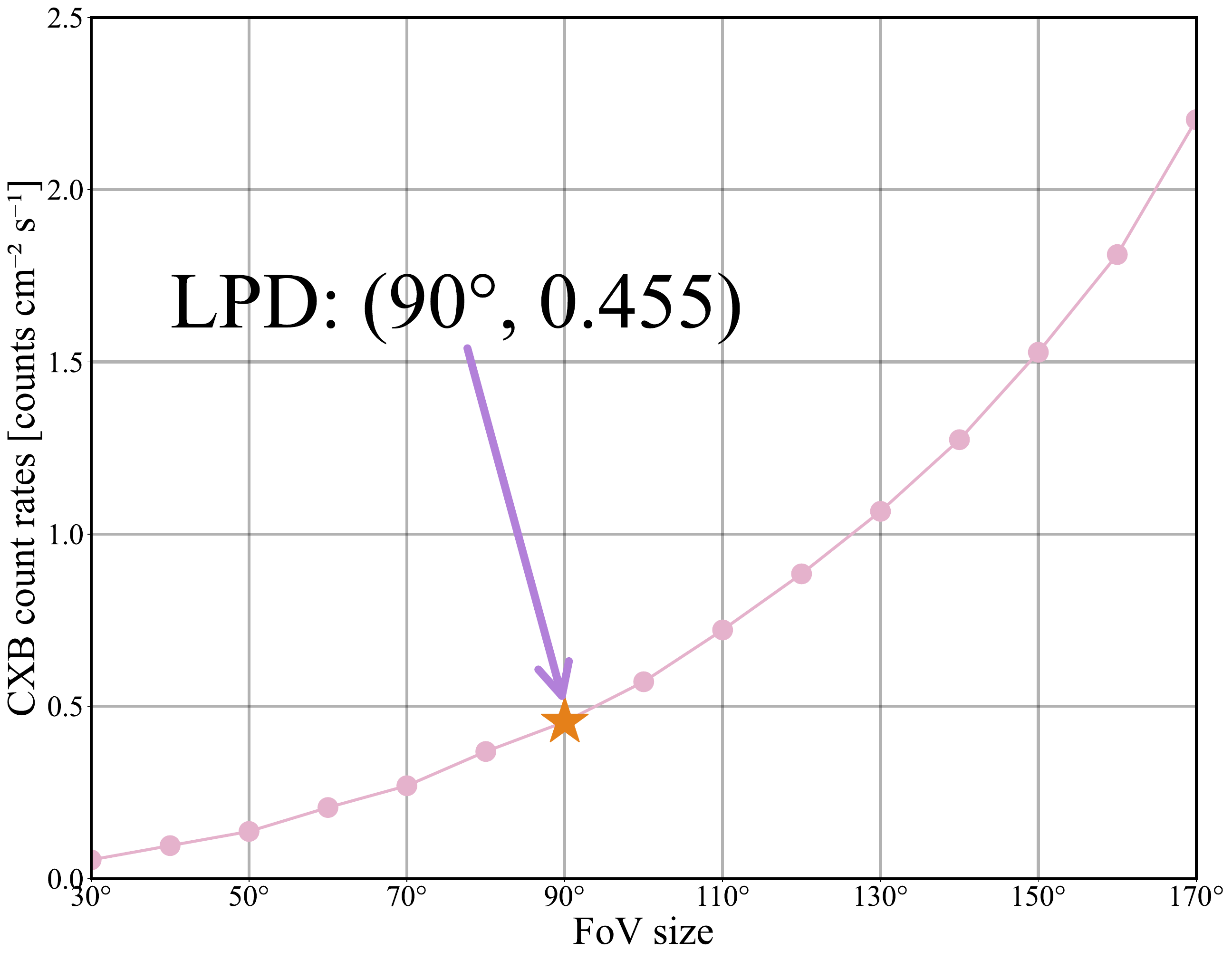}
\caption{The count rate of CXB changes with the FoV. At a 90° FoV, the count rate is 0.455 count $\mathrm{cm}^{-2}$$\mathrm{s}^{-1}$. Please note that in simulating the background response for different FoVs, the obstruction caused by the mechanical arm adapter was not considered. The simulation reveals that, with a 90-degree FoV, the mechanical arm adapter obstructs approximately 2\% of low-energy CXB photons, while the scattering of high-energy CXB photons can be deemed negligible.}
\label{CXB}
\end{figure}

The presence of the CSS during the simulation is time-consuming. We initially evaluated whether there is a significant influence from the CSS on the in-orbit background count rate. The simulation results are presented in Table \ref{spacestation}. The results of the simulations indicate that considering the presence of the CSS leads to varying degrees of reduction in the count rate of different charged particle backgrounds. This reduction is primarily attributed to the obstruction of charged particles by the CSS. Moreover, various shielding materials of the gas chamber, such as ceramics and Kovar alloy, have the capability to block secondary low-energy particles. This may explain why the background count rate does not increase when the presence of the CSS is taken into account. The outcomes also indicate that the count rate of photon background remains predominantly stable, given that photons primarily ingress from the FoV.
Due to the yet-to-be-determined relative positions of various payloads on the CSS, more detailed simulations will be undertaken at a later stage. Additionally, considering the minor impact of charged particles on our results (see Section \ref{sect: Background rejection}), the subsequent detailed simulations in this study will not consider the influence of the CSS. 

Fig. \ref{Gas_Edep} illustrates the simulation response of the primary and secondary charged particles, as well as the Cosmic X-ray Background (CXB) in the detector with a $90^{\circ}$ FoV. Table \ref{background_resp} provides the simulated responses in the detector for charged particles in the energy range of 2-10 keV at different magnetic latitudes. The results presented in the table indicate that as the geomagnetic latitude increases, the contribution of the primary background becomes more significant, while the overall background response remains relatively stable. These findings are consistent with the results reported in \citep{huang2021modeling}.

Furthermore, we conducted simulations to examine the background response for different FoVs. The results demonstrate that the simulated total count rate of charged particles does not change with the FoV of the detector. However, the total count rate of the CXB increases rapidly as the FoV expands within the range of 30 to 170 degrees, as depicted in Fig. \ref{CXB}. It is evident from the figure that the CXB constitutes the largest proportion of the background components under the $90^{\circ}$ FoV.

\begin{figure}
\centering
\includegraphics[scale=0.30]{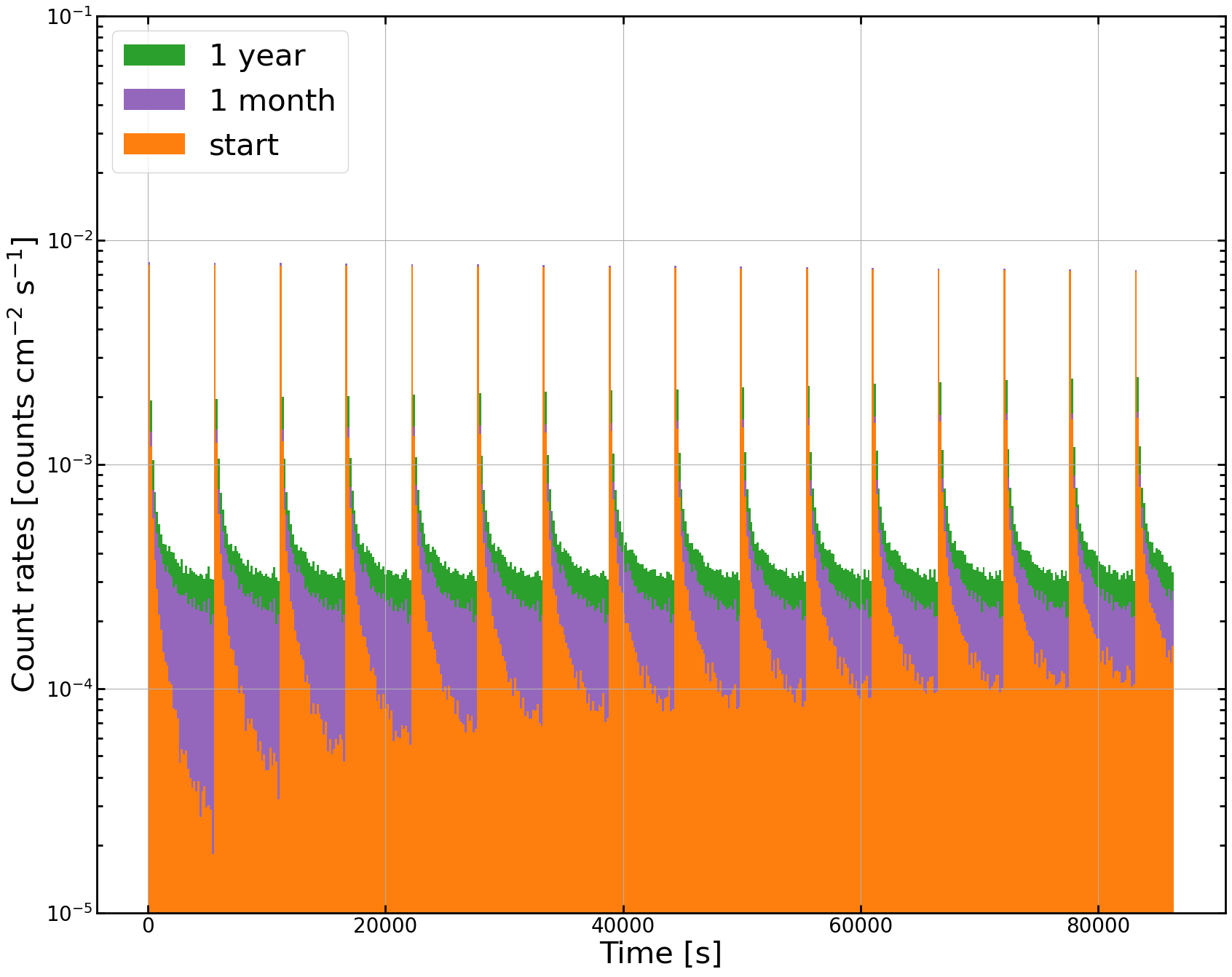}
\caption{The SAA delayed background changes with time within one day for three different timescales when LPD is in-orbit for beginning, 1 month and 1 year. Each bin in the graph represents 172.8 seconds. Due to the relatively long half-lives of certain elements, the background count rate experience a slight increase with the increase in mission duration.}
\label{timesaa}
\end{figure}

The delayed background caused by protons in the South Atlantic Anomaly (SAA) region decays rapidly once the LPD passes through this area. Fig. \ref{timesaa} illustrates the variation of SAA count rate within one day at different timescales: beginning, 1 month, and 1 year after the LPD is in-orbit. The purpose of this investigation is to examine the cumulative effect of the delayed background. The results indicate that the delayed background in the SAA region can be considered negligible compared to photon background. This may be due to the shielding effect provided by the gas chamber shell. The background primarily originates from materials within the chamber.

\begin{figure*}
\centering
\includegraphics[scale=0.8]{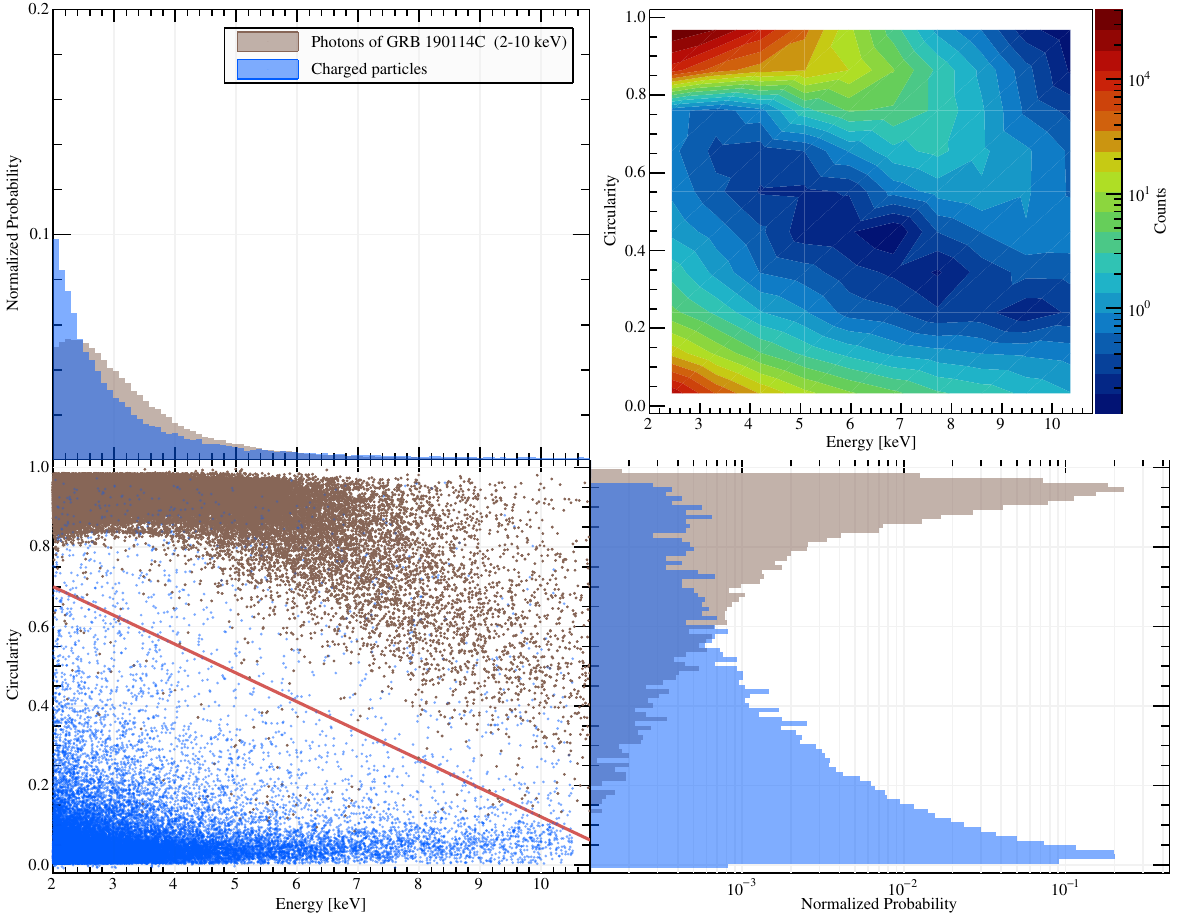}
\caption{The algorithm of charged particle background screening. The equation of the red straight line in the figure is y = 0.06x + 0.7. The background rejection rate after linear cut is 93.22\%, and the photon retention rate is 99.94\%. The total count rate of charged particle background before the cut is 0.43 $\operatorname{counts} / \mathrm{cm}^{2} / \mathrm{s}$ on average, and after the cut, it reduces to 0.03 $\operatorname{counts} / \mathrm{cm}^{2} / \mathrm{s}$.}
\label{Background_rejection}
\end{figure*}
\subsection{Background rejection for charged particles}

\label{sect: Background rejection}

% Due to the unknown incident direction of X-ray photoelectron tracks, distinguishing X-ray background from GRB X-ray photons poses a challenge. However, the morphological and characteristic parameter distributions of charged particle backgrounds significantly differ from those of X-ray photons. Therefore, background subtraction algorithms can be developed to reject charged particle backgrounds. Xie et al. conducted extensive research on the background of IXPE and compared various characteristic parameters of charged particle tracks with those of photoelectron tracks to achieve background rejection. Their methods achieved a background rejection efficiency of 75\%. In addition to these methods, comparing the circularity (which describes the degree of roundness of the track, ranging from 0 to 1, with values closer to 1 indicating a more circular track) and energy deposition distributions of charged particle tracks with photoelectron tracks can straightforwardly and efficiently reject the charged particle backgrounds, as shown in  Fig.  \ref{Background_rejection}. The rejection efficiency for charged particle background can exceed 90%.

Due to the unknown incident direction of X-ray photoelectron tracks, distinguishing X-ray background from GRB X-ray photons presents a challenge. However, there are noticeable differences in the morphological and characteristic parameter distributions between charged particle backgrounds and X-ray photons. This enables the development of background subtraction algorithms for rejecting charged particle backgrounds. Xie et al. \citep{xie2021study} conducted extensive research on the background of IXPE and compared various characteristic parameters of charged particle tracks with those of photoelectron tracks, resulting in a total background rejection efficiency of 75\%.

In addition to these methods, a straightforward and efficient approach for rejecting charged particle backgrounds is to compare the circularity \citep{vzunic2010hu,kitaguchi2018optimized} (a parameter describing the roundness of a track, ranging from 0 to 1, with values closer to 1 indicating a more circular track) and energy deposition distributions of charged particle tracks with those of photoelectron tracks. The rejection efficiency for charged particle backgrounds can exceed 90\%, as illustrated in Fig. \ref{Background_rejection}.

% \usepackage{booktabs}

\begin{figure*}
\centering
\includegraphics[scale=0.20]{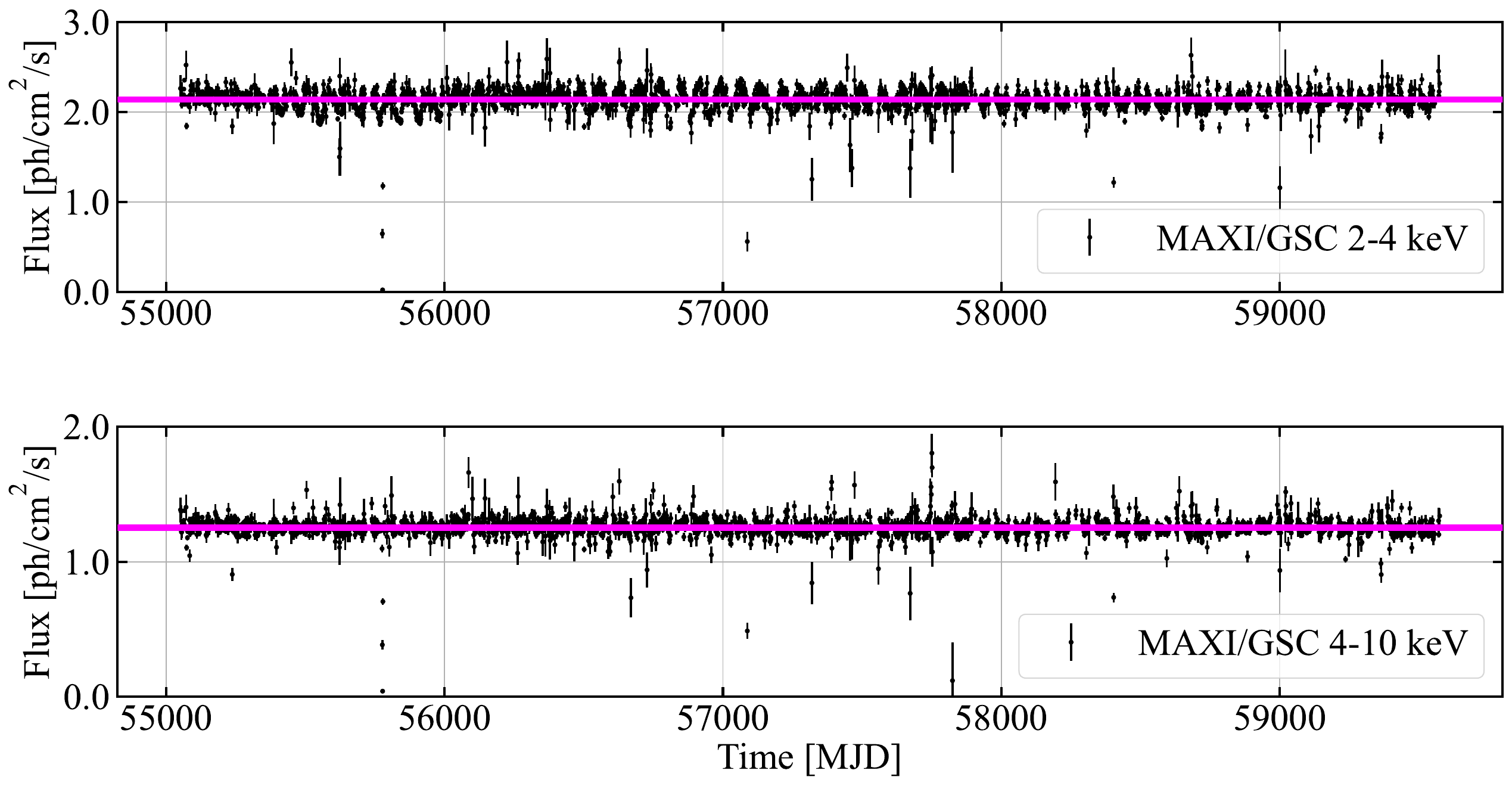}
\includegraphics[scale=0.20]{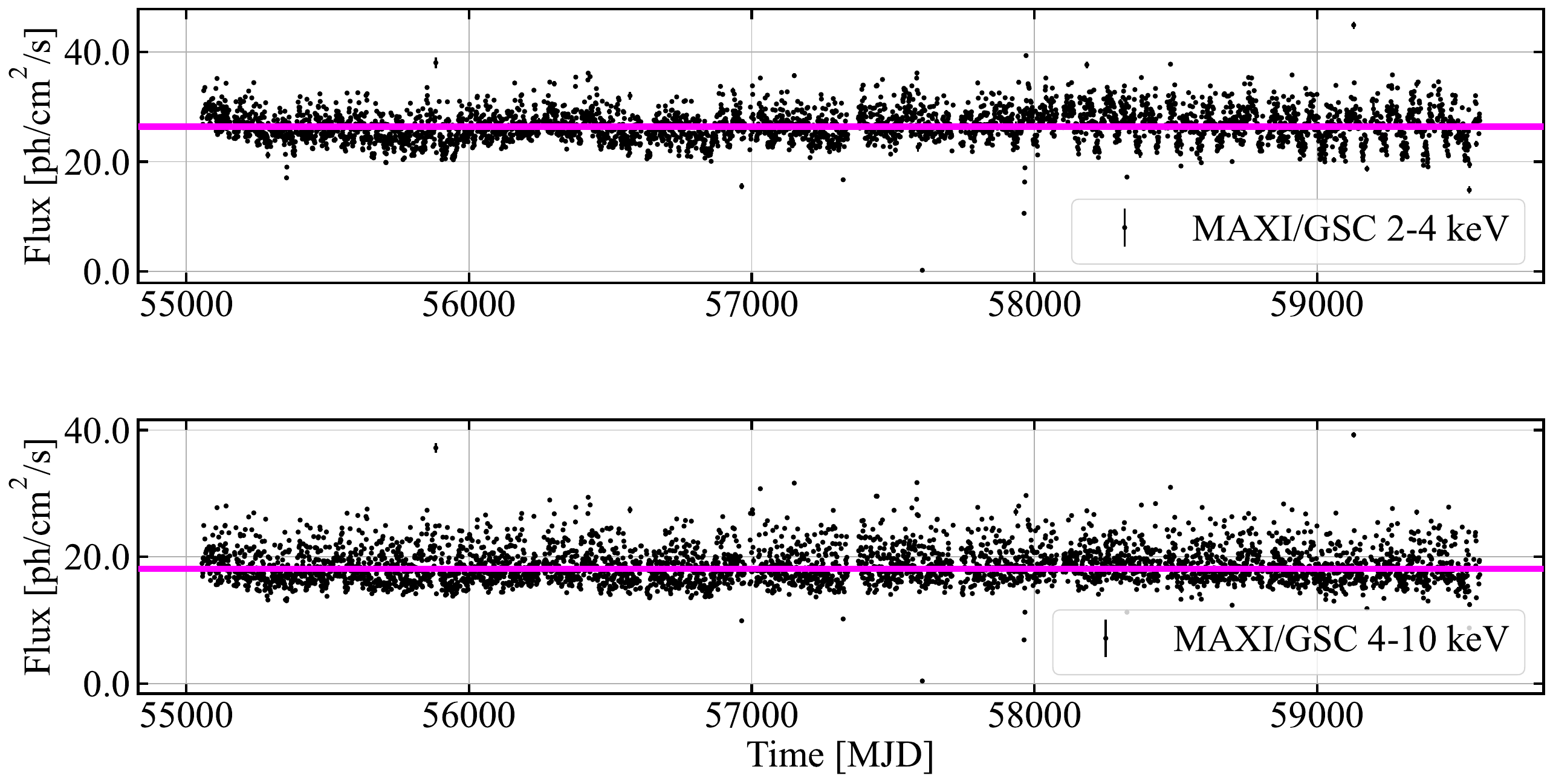}
\caption{Example of MAXI observation flux data respectively in 2-4 and 4-10 keV. Left panel: Crab; Right panel: Sco X-1. We take the median of the flux as the estimate value for the flux, as shown by the magenta line in the graph.}
\label{maxi}
\end{figure*}

\begin{figure*}
\centering
\includegraphics[scale=0.6]{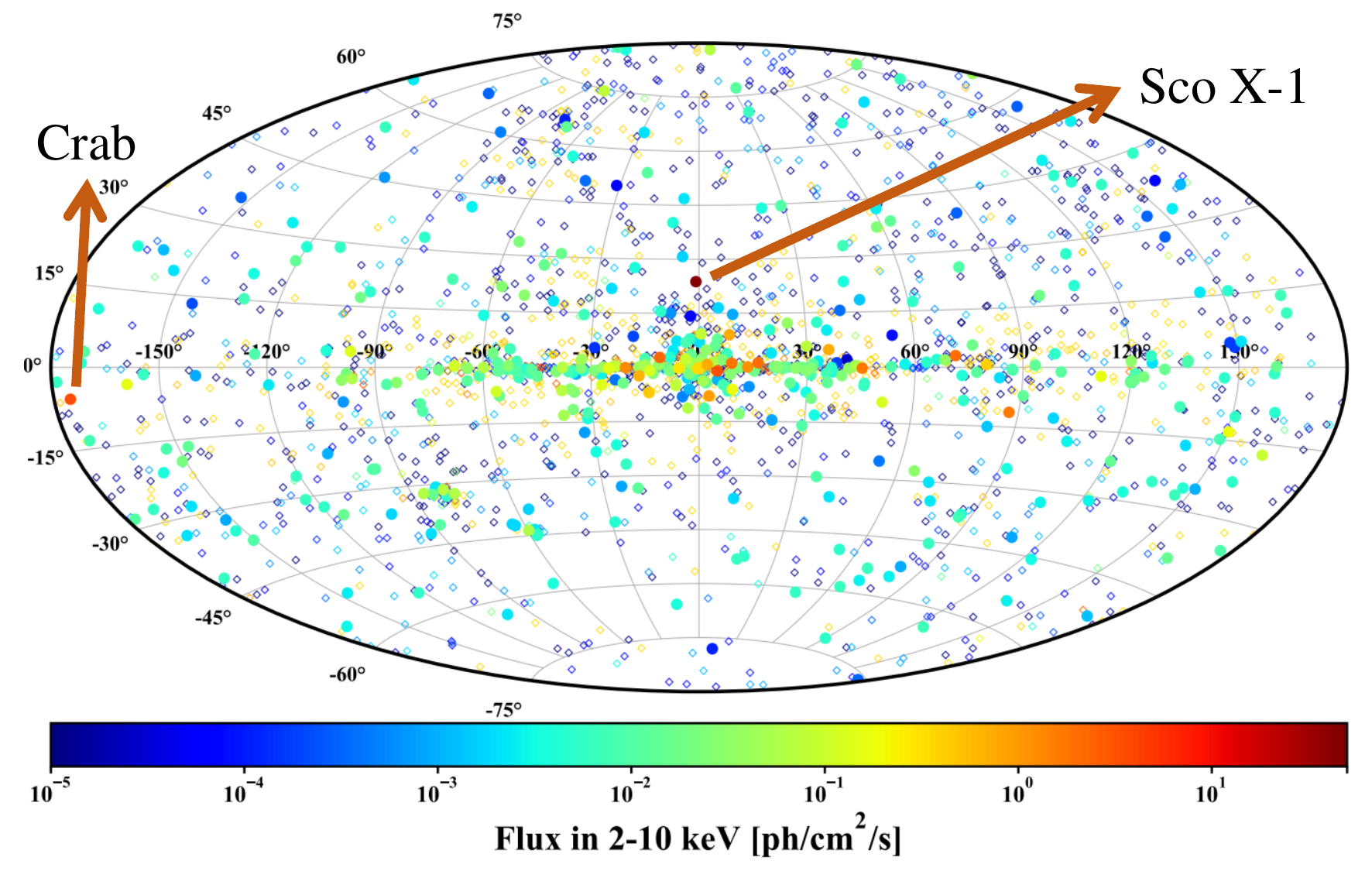}
\caption{Distribution of the sources on the sky in Galactic coordinates and the flux of each sources in 2-10 keV. The sources detected by MAXI/GSC are indicated by solid circles, and the sources detected just by Swift/BAT are indicated by hollow diamonds. The positions of Crab and Sco X-1 are indicated in the diagram with arrows.}
\label{source_catalogue}
\end{figure*}

\begin{figure}
\centering
\includegraphics[scale=0.42]{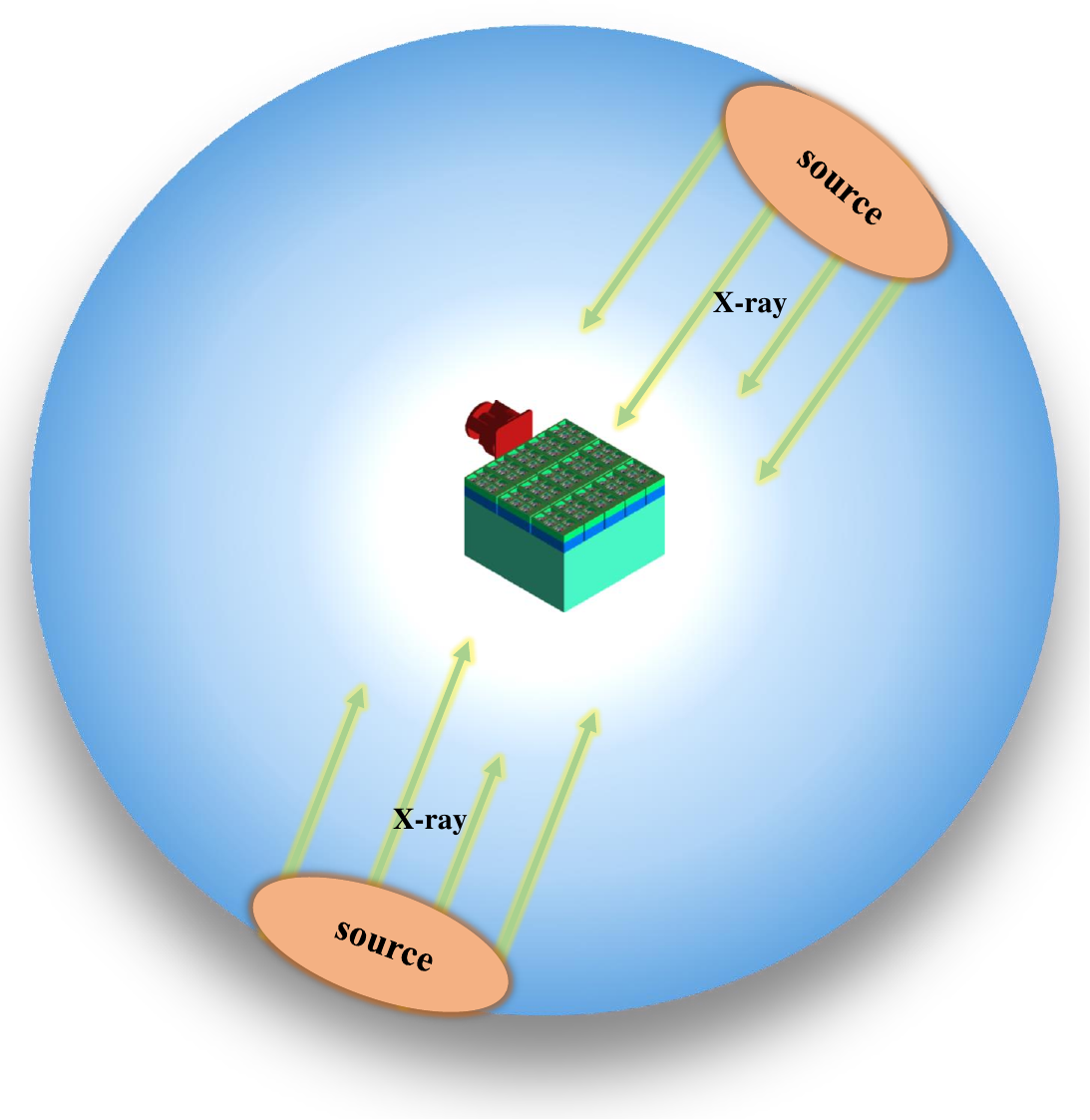}
\caption{
% Illustration of the setup for the X-ray source incident sources, with the LPD positioned at the center of the sphere. During the simulation of the X-ray sources, they are distributed on the surface of the sphere according to their position coordinate distribution, with emission directions pointing towards the center of the sphere. During the simulation of GRBs, their positions are randomly distributed throughout the sphere.
Illustration of the setup for the incident X-ray sources, with the LPD positioned at the center of the sphere. In the simulation of X-ray sources, they are distributed on the surface of the sphere based on their position coordinate distribution, with emission directions pointing towards the center of the sphere. 
}
\label{GRBsim}

\end{figure}

\begin{figure*}
\centering
\includegraphics[scale=0.33]{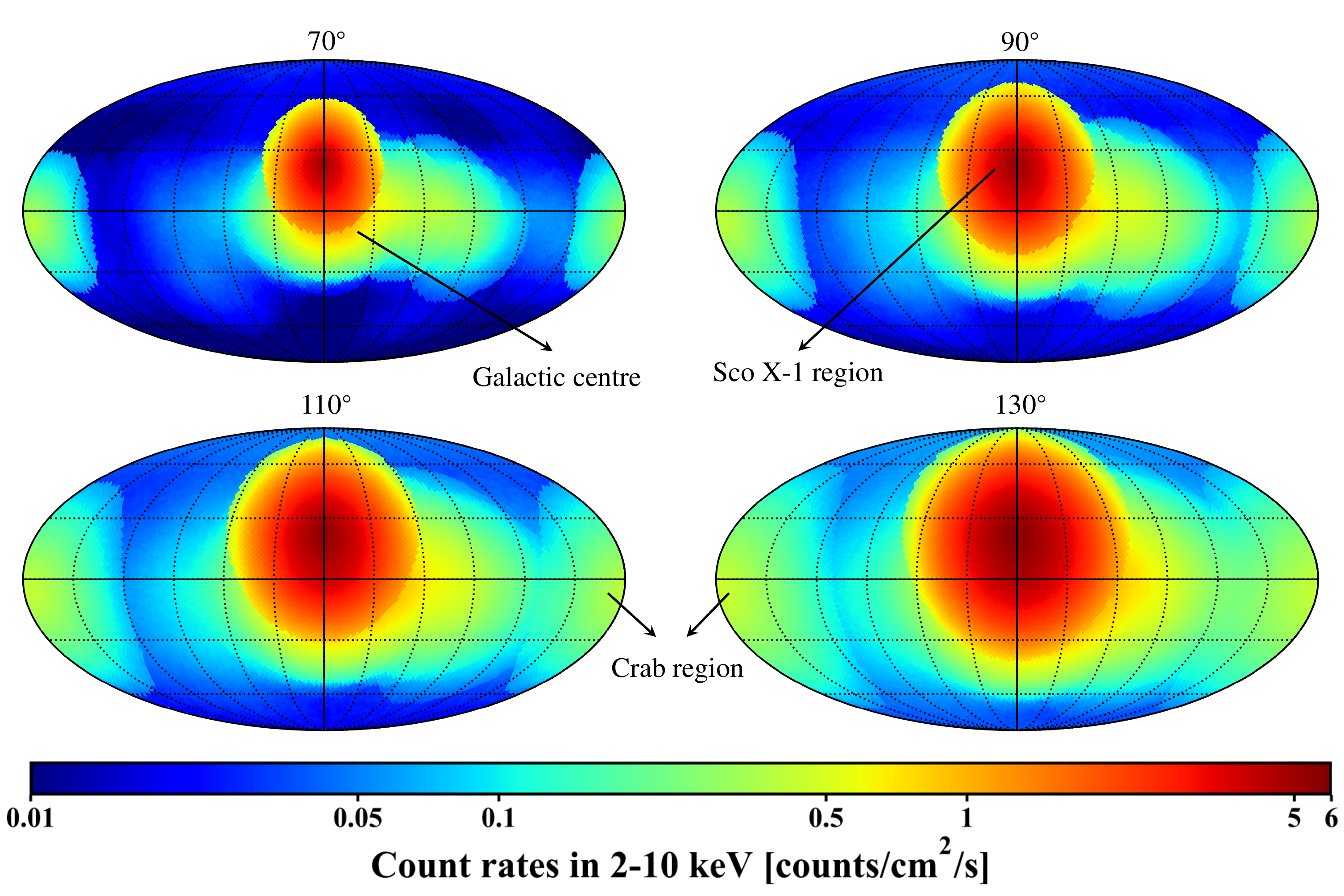}
\caption{Sky survey simulation maps. The count rate per square centimeter per second from all of the MAXI/GSC sources is depicted based on the pointing direction in the sky in Galactic coordinates (with longitude increasing from left to right). The four maps correspond to the four different FoVs: $70^{\circ}$, $90^{\circ}$, $110^{\circ}$, and $130^{\circ}$ (total 12288 pixels).}
\label{Simulated_map}
\end{figure*}

\begin{figure*}
\centering
\includegraphics[scale=0.3]{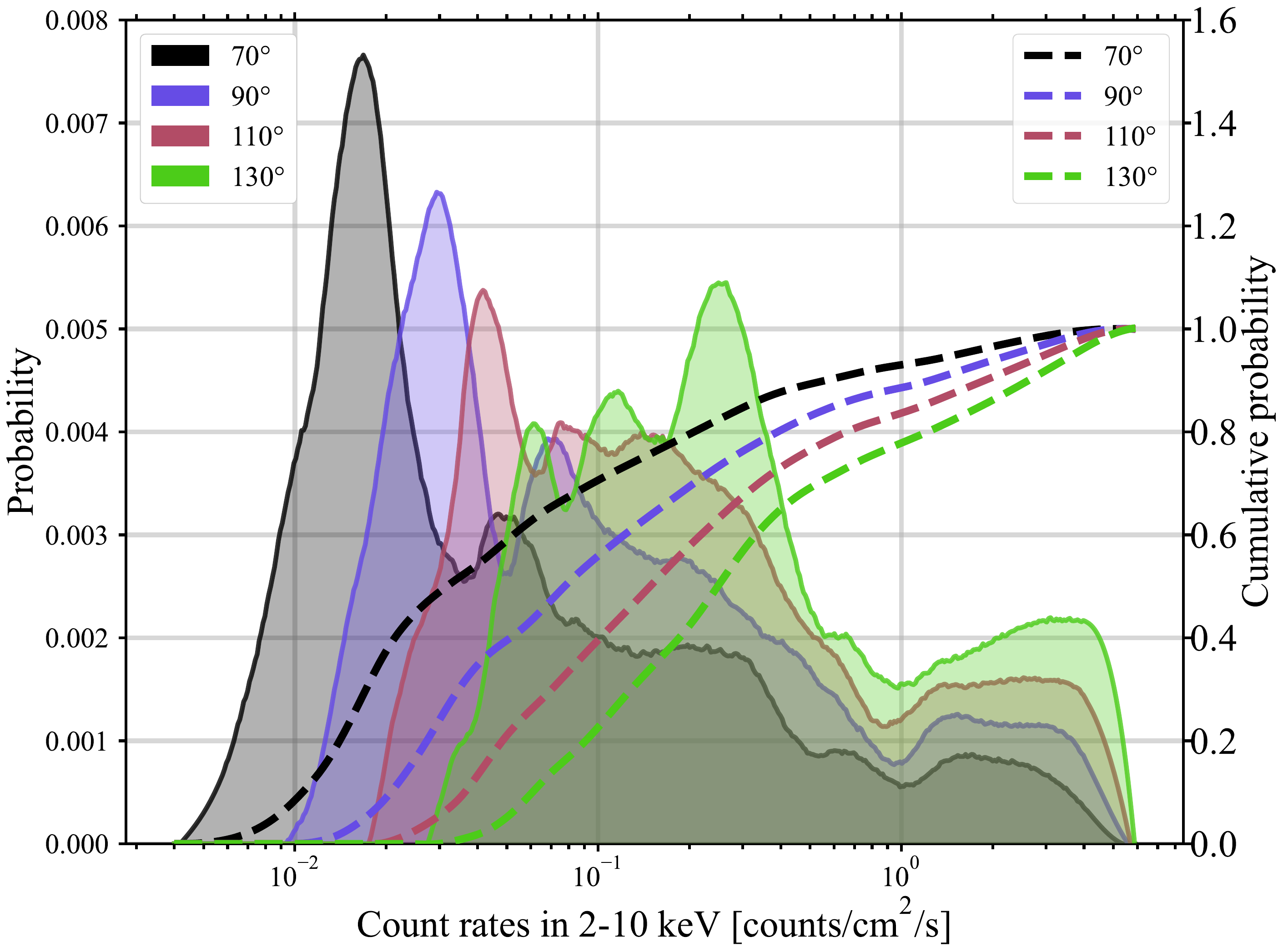}
\caption{Count rate distributions of sky survey simulation maps under different detector orientations and at four FoVs respectively are 70 degrees, 90 degrees, 110 degrees, and 130 degrees.}
\label{source}
\end{figure*}

\section{Catalogue of X-ray sources and simulated sky survey}
\label{sect:Src}
\subsection{Catalogue of X-ray sources}

To assess the influence of X-ray sources on gamma-ray burst polarization observations, it is essential to have information regarding the positions, flux, and energy spectra characteristics of each bright X-ray source distributed in space.
% MAXI/GSC \citep{matsuoka2009maxi} conducted all-day X-ray sky surveys and updates the observational data of X-ray sources on its official website daily. The data includes the positional coordinates of the X-ray sources and the flux variation over time in the energy ranges of 2-4, 4-10, and 10-20 keV. The example of flux data of Crab and Sco X-1 are shown in Fig. \ref{maxi}. The red line in the figure refers to the median flux. Dagoneau et al. have published Onboard catalogue of known X-ray sources for SVOM/ECLAIRs \citep{dagoneau2021onboard} in 2021. This article conduct a joint fitting of the flux observation data from MAXI/GSC and Swift/BAT (14-195 keV) \citep{barthelmy2005burst} using single powerlaw or broken powerlaw, and the best fitting spectral models and parameters for each X-ray source in MAXI/GSC and Swift/BAT are acquired. The fitting parameters for each source has been made public\footnote{\href{http:
% //cdsarc.u-strasbg.fr/viz-bin/cat/J/A+A/645/A18}{http:
% //cdsarc.u-strasbg.fr/viz-bin/cat/J/A+A/645/A18}}. From this source catalogue, we use all the MAXI/GSC sources and picked out the Swift/BAT sources that fit well at 2-10 keV. Fig. \ref{source_catalogue} shows the distribution of the sources on the sky in Galactic coordinates and the flux of each sources in 2-10 keV. It can be seen from the figure that the brightest X-ray sources in space in 2-10 keV are mainly distributed in the galactic plane except for Crab and Sco X-1.

MAXI/GSC \citep{matsuoka2009maxi} has conducted comprehensive X-ray sky surveys throughout the day and provides daily updates of observational data for X-ray sources on its official website\footnote{\href{http://maxi.riken.jp/top/slist.html}{http://maxi.riken.jp/top/slist.html}}. The data includes positional coordinates of the X-ray sources and flux variations over time in the energy ranges of 2-4, 4-10, and 10-20 keV. An example of flux data for Crab and Sco X-1 is shown in Fig. \ref{maxi}. We take the median of the flux as the estimate value for the flux, as shown by the magenta line in the graph.

In 2021, Dagoneau et al. published the Onboard catalogue of known X-ray sources for SVOM/ECLAIRs \citep{dagoneau2021onboard}. This article performs joint fitting of flux observation data from MAXI/GSC and Swift/BAT (14-195 keV) \citep{barthelmy2005burst} using single power law or broken power law models. The best-fitting spectral models and parameters for each X-ray source in MAXI/GSC and Swift/BAT are obtained. The fitting parameters for each source are publicly available\footnote{\href{http:
//cdsarc.u-strasbg.fr/viz-bin/cat/J/A+A/645/A18}{http:
//cdsarc.u-strasbg.fr/viz-bin/cat/J/A+A/645/A18}}. From this source catalogue, we utilize all the MAXI/GSC sources and select the Swift/BAT sources that fit well in the 2-10 keV energy range.
Fig. \ref{source_catalogue} shows the sky distribution of the sources in Galactic coordinates and their flux in the 2-10 keV range. It is evident from the figure that the brightest X-ray sources in space in the 2-10 keV range are predominantly located in the galactic plane, except for Crab and Sco X-1.

\subsection{Sky survey simulations}
% The input scripts for each MAXI/GSC sources are generated for simulate detections using Geant4. Taking into account computational resource constraints and the mismatch in observed energy ranges, the sources from the Swift/BAT catalog are not using for simulation. Each scripts include the initial position coordinates, emission direction, spectral, and total flux information of the sources. The incident sources is modeled as a circular planar source with a sufficiently large area to cover the entire detector as shown in Fig. \ref{GRBsim}. We change the pointing direction of the detector to realize the simulated all sky surveys of the X-ray source. We total changed the pointing direction of the detector 12,288 times to draw all sky survey maps. The simulated maps of bright sources in detectors with different detector pointing direction respectively in $70^{\circ} \times 70^{\circ}$, $90^{\circ} \times 90^{\circ}$, $110^{\circ} \times 110^{\circ}$, $130^{\circ} \times 130^{\circ}$ FoV size are shown in Fig. \ref{Simulated_map}.

Input scripts are generated for each MAXI/GSC source to simulate detections using Geant4. Due to computational resource limitations and the mismatch in observed energy ranges, the sources from the Swift/BAT catalog are not used for simulation. Each script includes the initial position coordinates, emission direction, spectral information, and total flux of the sources. The incident sources are modeled as circular planar sources with sufficiently large areas to cover the entire detector, as illustrated in Fig. \ref{GRBsim}.

To achieve simulated all-sky surveys of X-ray sources, we vary the pointing direction of the detector. We changed the pointing direction of the detector a total of 12,288 times to generate all-sky survey maps. The simulated maps of bright sources in detectors with different pointing directions, covering FoVs of $70^{\circ}$, $90^{\circ}$, $110^{\circ}$, and $130^{\circ}$, are presented in Fig. \ref{Simulated_map}. The count rate distributions of X-ray sources for different detector orientations are presented in Fig. \ref{source} across the four FoVs.

\section{Summary and discussion}
\label{sect:summary}

This paper presents a background simulation study for a new scheme to measure the polarization of GRB prompt emission in soft X-ray range using a gas photoelectric polarization detector with a wide FoV. 

Comparing our background simulation results, the CXB constitutes the largest proportion among various background components. The total background count rate is $\sim$ 0.55 count$/ \mathrm{cm}^2 / \mathrm{s}$ after charged particle rejection at $90^{\circ}$ FoV. As the FoV expands, the count rate of the CXB experiences a rapid increase within the FoV range of 30 to 170 degrees. and the probability of bright X-ray sources in the galactic plane as well as the brightest X-ray source, Sco X-1, Crab, entering the FoV increases. In addition, as the FoV expands, the probability of the Sun entering the FoV also increases. The probability of the Sun entering the FoV depends on the specific orbit and the FoV of the detector. For example, for the LPD orbit and a FoV of 90 degrees, the probability of the Sun entering the FoV in a year is approximately 1/5. According to the GOSE's solar X-ray monitoring data\footnote{\href{https://solarmonitor.org/}{https://solarmonitor.org/}}, the solar flux in the 2-10 keV range fluctuates between several thousands to tens of thousands photons$/ \mathrm{cm}^2 / \mathrm{s}$ \citep{10.1117/12.254078}. Currently, in the laboratory, the LPD detector is capable of tolerating a count rate of around 5000 counts$/ \mathrm{cm}^2 / \mathrm{s}$. Observing the Sun is also possible. However, the Sun's surface is quite active, resulting in significant count rate variations. Additionally, the differences in solar flux across different years are substantial. Hence, further research is needed to determine whether it is necessary to power on during the Sun's presence in the FoV. The ability to withstand solar observations is our ultimate design goal due to its immense scientific potential. Recent observations by IXPE and PolarLight of brighter sources, such as Sco X-1, have provided high-confidence polarization data. PolarLight observed a flux-dependent low polarization in the 4-8 keV range and no polarization in the 3-4 keV range for Sco X-1 \citep{Long_2022}. Therefore, when these bright sources are located at the edge of the FoV, it is still possible to observe bright gamma-ray bursts with high confidence.
% This increase in the CXB background response and the presence of bright X-ray sources entering the FoV can result in a reduction in the effective observation time of the gas photoelectric polarization detector or a decrease in the signal-to-noise ratio.

When determining the FoV of the detector, it is crucial to consider not only the background level under different FoVs but also to combine GRB observation simulations (to be discussed in future work) with the expected orbit or effective observation time of the mission.

From the simulation results, it is evident that a wide FoV significantly increases the photon background level of the detector, which is unavoidable when detecting transient sources. However, the flux of most GRB prompt emissions is sufficiently high in soft X-ray range. For example, for GRB190114C, using the spectral data provided on the Swift official website, the flux in the 2-10 keV range is estimated to be approximately 15 photons$/ \mathrm{cm}^2 / \mathrm{s}$. With a weight modulation factor of 0.33 and an average background level of 0.55 count$/ \mathrm{cm}^2 / \mathrm{s}$, the calculated sensitivity using Eq. (\ref{eq:MDP}) for a 20-degree oblique incidence is approximately 7.8\% in a 300-second observation.
Nonetheless, in order to enhance the detector's performance and achieve greater scientific output, it remains an important research topic for us to optimize the detector and minimize the in-orbit background in the future.

\section*{Acknowledgements}
This work is supported by Department of Physics and GXU-NAOC Center for Astrophysics and Space Sciences, Guangxi University. This work is supported by the National Natural Science Foundation of China (Grant Nos. 12027803, U1731239, 12133003, 12175241, U1938201, U1732266), the Guangxi Science Foundation (Grant Nos. 2018GXNSFGA281007, 2018JJA110048).

% \section*{DATA AVAILABILITY}
% The data underlying this article are available in the articl.

\bibliographystyle{apj}
\bibliography{ref.bib}

\end{document}